\documentclass[12pt]{article}
\usepackage{amsmath, amssymb}
\usepackage{graphicx}
\usepackage{natbib}
\usepackage{url} 
\usepackage{color, colortbl}
\usepackage{geometry}
\usepackage{array}
\usepackage{caption}
\usepackage{subcaption}
\usepackage{setspace}
\usepackage{hyperref}
\hypersetup{
  colorlinks= true,
  linkcolor= cyan,
  citecolor= blue,
}

\newcommand{\xbf}{\boldsymbol{x}}
\newcommand{\Xbf}{\boldsymbol{X}}
\newcommand{\zbf}{\boldsymbol{z}}
\newcommand{\Zbf}{\boldsymbol{Z}}
\newcommand{\bbeta}{\boldsymbol{\beta}}
\newcommand{\bgamma}{\boldsymbol{\gamma}}
\newcommand{\btheta}{\boldsymbol{\theta}}
\newcommand{\bxi}{\boldsymbol{\xi}}

\definecolor{lightgray}{gray}{0.9} 

\begin{document}

\title{\hspace{1.2cm} \bf Laplacian-P-splines for Bayesian \\
\hspace{1.2cm} inference in the mixture cure model}

\author{Oswaldo Gressani $^{a,}$\thanks{Corresponding author. E-mail address: \textit{$oswaldo.gressani@uhasselt.be$}} \ ,Christel Faes$^{a}$\ , Niel Hens$^{a,b}$\
\\
$^a$ Interuniversity Institute
for Biostatistics and statistical Bioinformatics (I-BioStat), \\
Data Science Institute, Hasselt University, Belgium\\
$^b$ Centre for Health Economics Research and Modelling Infectious Diseases, \\ Vaxinfectio, University of Antwerp, Belgium\\
}	

\date{}	
\maketitle	

\begin{center}
\textbf{Abstract}
\end{center}

\noindent The mixture cure model for analyzing survival data is characterized by the assumption that the population under study is divided into a group of subjects who will experience the event of interest over some finite time horizon and another group of cured subjects who will never experience the event irrespective of the duration of follow-up. When using the Bayesian paradigm for inference in survival models with a cure fraction, it is common practice to rely on Markov chain Monte Carlo (MCMC) methods to sample from posterior distributions. Although computationally feasible, the iterative nature of MCMC often implies long sampling times to explore the target space with chains that may suffer from slow convergence and poor mixing. Furthermore, extra efforts have to be invested in diagnostic checks to monitor the reliability of the generated posterior samples. An alternative strategy for fast and flexible sampling-free Bayesian inference in the  mixture cure model is suggested in this paper by combining Laplace approximations and penalized B-splines. A logistic regression model is assumed for the cure proportion and a Cox proportional hazards model with a P-spline approximated baseline hazard is used to specify the conditional survival function of susceptible subjects. Laplace approximations to the conditional latent vector are based on analytical formulas for the gradient and Hessian of the log-likelihood, resulting in a substantial speed-up in approximating posterior distributions. The spline specification yields smooth estimates of survival curves and functions of latent variables together with their associated credible interval are estimated in seconds. The statistical performance and computational efficiency of the proposed Laplacian-P-splines mixture cure (LPSMC) model is assessed in a simulation study. Results show that LPSMC is an appealing alternative to classic MCMC for approximate Bayesian inference in standard mixture cure models. Finally, the novel LPSMC approach is illustrated on three applications involving real survival data.\\


\section{Introduction}

\noindent Survival analysis is a challenging, yet very attractive area of statistical science that is devoted to the study of time-to-event data. Standard models for survival data typically leave no room for the existence of a cure fraction such that it is implicitly assumed that all subjects of the population under study will experience the event of interest as time unfolds for a sufficiently long period. Technological breakthroughs in medicine during the last decades, especially in cancer research, have led to the development of promising new treatments and therapies so that many diseases previously considered fatal can now be cured. This phenomenon has triggered the necessity to develop models that allow for long-term survivors and gave birth to cure models. A recent complete textbook treatment on cure models is proposed by \cite{peng2020cure} and an enriching literature review on cure regression models has been written by \cite{amico2018cure}. Among the large family of cure models that have emerged, the mixture cure model driven by the seminal work of \cite{boag1949maximum,berkson1952survival,haybittle1965two} and later refined by \cite{farewell1977combined,farewell1982use} is probably the most prominent as its mathematical formulation allows for a clear and interpretable separation of the population in two categories, namely cured subjects and susceptible subjects who are at risk of experiencing the event of interest.\\
\indent Let $T \in [0,+\infty)$ be a continuous random variable representing the survival time. Existence of a cure proportion in the population under study is made possible by allowing the event $\{T=+\infty\}$ to arise with positive probability. To include covariate information, denote by $\Xbf$ and $\Zbf$ random covariate vectors (with continuous and/or discrete entries) that belong to covariate spaces $\mathcal{X}$ and $\mathcal{Z}$, respectively. In a mixture cure model, the population survival function expresses the separation between the cured and uncured subpopulations as follows:

\vspace{-1.4cm} 

\begin{eqnarray} \label{Spop}
S_p(t \vert \xbf, \zbf)=1-p(\xbf)+p(\xbf) S_u(t \vert \zbf),
\end{eqnarray}

\noindent with covariate vectors $\xbf=(x_1,\dots,x_p)^{\top} \in \mathcal{X}$  and $\zbf=(z_1,\dots,z_q)^{\top} \in \mathcal{Z}$ that can share (partially) the same components or can be entirely different. The term $p(\xbf)$ is frequently called the ``incidence'' of the model and corresponds to the conditional probability of being uncured, i.e.\ $p(\xbf)=P(B=1\vert \Xbf=\xbf)$ with binary variable $B=\mathbb{I}(T<+\infty)$ referring to the (unknown) susceptible status and $\mathbb{I}(\cdot)$ the indicator function, i.e.\ $\mathbb{I}(E)=1$ if condition $E$ is true. The term $S_u(t\vert \zbf)$ is known as the ``latency'' and represents the conditional survival function of the uncured subjects $S_u(t \vert \zbf)=P(T>t \vert B=1, \Zbf=\zbf)$. The logistic link is commonly employed to establish a functional relationship between the probability to be uncured and the vector $\xbf$ \citep{farewell1977model,ghitany1994exponential,taylor1995semi}, so that $p(\xbf)=\exp(\beta_0+\xbf^{\top}\bbeta)/(1+\exp(\beta_0+\xbf^{\top}\bbeta))$, with regression coefficients $\bbeta=(\beta_1,\dots,\beta_p)^{\top}$ and $\beta_0$ an intercept term. The latency part is often specified in a semiparametric fashion by using the \cite{cox1972regression} proportional hazards (PH) model \citep[see e.g.][]{kuk1992mixture,peng2000nonparametric} and implies the following form for the survival function of the susceptibles $S_u(t \vert \zbf)=S_0(t)^{\exp(\zbf^{\top} \bgamma)}$, where $\bgamma=(\gamma_1,\dots,\gamma_q)^{\top}$ are the regression coefficients pertaining to the latency part and $S_0(\cdot)$ is the baseline survival function.\\
\indent The philosophy underlying Bayesian approaches considers that the model parameters are random and that their underlying uncertainty is characterized by probability distributions. After obtaining data, Bayes' theorem acts as a mechanistic process describing how to update our knowledge and is essentially the key ingredient permitting the transition from prior to posterior beliefs. Unfortunately, the complexity of mixture cure models is such that the posterior distribution of latent variables of interest are not obtainable in closed form. An elegant stochastic method that is widely used in practice is Markov chain Monte Carlo (MCMC) as it allows to draw random samples from desired target posterior distributions and hence compute informative summary statistics. According to \cite{greenhouse1996practical}, the first steps of Bayesian methods applied to mixture models with a cure fraction date back as far as \cite{chen1985bayesian} to analyze cancer data. Later, \cite{stangl1991modeling} and \cite{stangl1998assessing} used a Bayesian mixture survival model to analyze clinical trial data related to mental health. The end of 1990s saw the emergence of Bayesian approaches in the promotion time cure model \citep{chen1999new}, another family of cure models motivated by biological mechanisms that does not impose a mixture structure on the survival. Some references for Bayesian analysis in the latter model class are \cite{yin2005cure}, who proposed a Box-Cox based transformation on the population survival function to reach a unified family of cure rate models embedding the promotion time cure model as a special case; \cite{bremhorst2016flexible} used Bayesian P-splines with MCMC for flexible estimation in the promotion time cure model and \cite{gressani2018fast} suggested a faster alternative based on Laplace approximations. More recent uses of Bayesian methods in mixture cure models are \cite{yu2012bayesian} in the context of grouped population-based cancer survival data or \cite{martinez2013mixture} who consider a parametric specification for the baseline survival of uncured subjects governed by a generalized modified Weibull distribution. In the literature of cure survival models, only scarce attempts have been initiated to propose an alternative to the deep-rooted MCMC instruments. This is especially true for mixture cure models, where to our knowledge \cite{lazaro2020approximate} is the only reference proposing an approximate Bayesian method based on a combination of Integrated Nested Laplace Approximations (INLA) \citep{rue2009approximate} and modal Gibbs sampling \citep{gomezrubio2017mixture}. In this article, we propose a new approach for fast approximate Bayesian inference in the mixture cure model based on the idea of Laplacian-P-splines \citep{gressani2018fast}. The proposed Laplacian-P-splines mixture cure (LPSMC) model has various practical and numerical advantages that are worth mentioning. First, as opposed to \cite{lazaro2020approximate}, our approach is completely sampling-free in the sense that estimation can be fully reached without the need of drawing samples from posterior distributions. This of course implies a huge gain from the computational side, without even mentioning the additional speed-up effect implied by the analytically available gradient and Hessian of the log-likelihood function in our Laplace approximation scheme. Second, the LPSMC approach delivers approximations to the joint posterior latent vector, while the INLA scheme concentrates on obtaining approximated versions of the marginal posterior of latent variables. A direct positive consequence is that with LPSMC, the ``delta'' method can be used to compute (approximate) credible intervals for functions of latent variables, such as the cure proportion or the survival function of the uncured, in virtually no time. A third beneficial argument is that the use of P-splines is particularly well adapted in a Bayesian framework and provide smooth estimates of the survival function. Finally, our approach and its associated algorithms are explicitly constructed to fit mixture cure models contrary to INLA that cannot fit such models originally \citep{lazaro2020approximate}.\\
\indent The article is organized as follows. In Section 2, the spline specification of the log-baseline hazard is presented and the Bayesian model is formulated along with the prior assumptions. Laplace approximations to the conditional latent vector are derived and an approximate version of the posterior penalty parameter is proposed. The end of Section 2 is dedicated to the construction of approximate credible intervals for (functions of) latent variables. Section 3 aims at assessing the proposed LPSMC methodology in a numerical study with simulated data under different cure and censoring scenarios. Section 4 is dedicated to three real data applications and Section 5 concludes the article.

\section{The Laplacian-P-spline mixture cure model} 

\noindent We consider that the survival time $T$ is accompanied by the frequently encountered feature of random right censoring. Rather than observing $T$ directly, one observes the pair $(T_{\text{obs}},\tau)$, where $T_{\text{obs}}=\min(T,C)$ is the follow-up time and $\tau=I(T \leq C)$ is the event indicator ($\tau=1$ if the event occurred and $\tau=0$ otherwise) and $C$ is a non-negative random censoring time that is assumed conditionally independent of $T$ given the covariates, i.e.\ $C \perp T \vert \Xbf, \Zbf$. At the sample level, $\mathcal{D}_i =(t_i, \tau_i, \xbf_i, \zbf_i)$ denotes the observables for the $i$th unit, with $t_i$ the realization of $T_{\text{obs}}$ and $\tau_i$ its associated event indicator. Vectors $\xbf_i$ and $\zbf_i$ represent the observed covariate values of subject $i$ and the entire information set available from data with sample size $n$ is denoted by $\mathcal{D}=\bigcup_{i=1}^n$.

\subsection{Flexible modeling of the baseline risk function with B-splines}

\noindent A flexible spline specification of the (log) baseline hazard function $h_0(\cdot)$ is proposed \citep{whittemore1986survival,rosenberg1995hazard} using a linear combination of cubic B-splines, i.e.\ $\log h_0(t)=\btheta^{\top} b(t)$, where $\btheta=(\theta_1,\dots,\theta_K)^{\top}$ is a $K-$dimensional vector of B-spline amplitudes and $b(\cdot)=(b_1(\cdot),\dots,b_K(\cdot))^{\top}$ is a cubic B-spline basis constructed from a grid of equally spaced knots in the closed interval $\mathcal{I}=[0,t_u]$, with $t_u$ the largest observed follow-up time. Partitioning $\mathcal{I}$ into $J$ (say 300) sections of equal length $\Delta$ with midpoint $s_j$, the Riemann midpoint rule is used to approximate the analytically unsolvable baseline survival function:

\vspace{-1cm} 

\begin{eqnarray}
S_0(t)&=&\exp\left(-\int_0^t \exp\Big(\btheta^{\top} b(s)\Big) ds \right) \nonumber \\
&\approx& \exp\left(-\sum_{j=1}^{j(t)} \exp\Big(\btheta^{\top} b(s_j)\Big) \Delta \right), \nonumber 
\end{eqnarray}

\noindent where $j(t) \in \{1,\dots,J\}$ is an integer enumerating the interval that includes time point $t$.

\subsection{Latent vector and priors}

\noindent The latent vector of the model is $\bxi=(\btheta^{\top},\breve{\bbeta}^{\top},\bgamma^{\top})^{\top}$ and contains the spline vector $\btheta$, the vector of regression coefficients belonging to the incidence part (including the intercept) $\breve{\bbeta}=(\beta_0,\bbeta^{\top})^{\top}$ and the vector of remaining regression coefficients belonging to the latency part 
$\bgamma$, with dimension $\text{dim}(\bxi)=K+(p+1)+q$. Based on the idea of \cite{eilers1996flexible}, we fix $K$ large enough to ensure flexible modeling of the baseline hazard curve and counterbalance the latter flexibility by imposing a discrete penalty on neighboring B-spline coefficients based on finite differences. In a Bayesian translation \citep{lang2004bayesian}, the prior distributional assumption on the B-spline vector is taken to be Gaussian  $\btheta \vert \lambda \sim \mathcal{N}_{\text{dim}(\btheta)}\big(0,(\lambda P)^{-1}\big)$, with a covariance matrix formed by the product of a roughness penalty parameter $\lambda>0$ and a penalty matrix $P=D_r^{\top}D_r+\varepsilon I_K$ obtained from $r$th order difference matrices $D_r$ of dimension $(K-r)\times K$. An $\varepsilon$-multiple of the $K-$dimensional identity matrix $I_K$ is added to ensure full rankedness \citep{lambert2011smooth} with typical values for the scalar perturbation being $\varepsilon=10^{-6}$ \citep{eilers2021practical} or $\varepsilon=10^{-5}$ \citep{alston2013case}. Furthermore, a Gaussian prior is imposed on the remaining regression coefficients, with zero mean and small (common) precision $\zeta=10^{-6}$, resulting in the following proper (conditional) prior for the latent vector $\bxi \vert \lambda \sim \mathcal{N}_{\text{dim}(\bxi)}(0,\Sigma_{\bxi}(\lambda))$ with covariance matrix:

\vspace{-0.7cm} 

\begin{eqnarray}
\Sigma_{\bxi}(\lambda)=\begin{pmatrix}
(\lambda P)^{-1} & 0 \\
0 & \zeta^{-1}I_{(p+1)+q}
\end{pmatrix}. \nonumber 
\end{eqnarray}

\vspace{0.3cm}

\noindent  The prior precision matrix of the latent vector is denoted by $Q_{\bxi}(\lambda)=\Sigma^{-1}_{\bxi}(\lambda)$. For full Bayesian treatment, we impose a Gamma prior with mean $a_{\lambda}/b_{\lambda}$ and variance $a_{\lambda}/b_{\lambda}^2$ on the roughness penalty parameter $\lambda \sim \mathcal{G}(a_{\lambda},b_{\lambda})$. Fixing $a_{\lambda}=1$ and $b_{\lambda}=10^{-5}$ \citep[see e.g.][]{ccetinyurek2011smooth} yields a large variance and hence reflects a minimally informative prior for $\lambda$. Other prior specifications are also available \citep[see e.g.][]{jullion2007robust,ventrucci2016penalized}.

\subsection{Laplace approximations}

\noindent In a mixture cure model, the full likelihood is given by \citep[see e.g.][]{sy2000estimation}:

\vspace{-1cm}

\begin{eqnarray}
\mathcal{L}(\bxi; \mathcal{D})=\prod_{i=1}^n \Big(p(\xbf_i) f_u(t_i \vert \zbf_i)\Big)^{\tau_i} \Big(1-p(\xbf_i)+p(\xbf_i)S_u(t_i\vert \zbf_i)\Big)^{(1-\tau_i)}, \nonumber 
\end{eqnarray}

\noindent where $f_u(t_i \vert \zbf_i)=-(d/dt) S_u(t_i \vert \zbf_i)$. Using the Cox PH model specification for the survival function of the susceptibles, one recovers:

\vspace{-1cm} 

\begin{eqnarray}
f_u(t_i \vert \zbf_i)&=&-(d/dt) S_u(t_i \vert \zbf_i) \nonumber \\
&=& -(d/dt) S_0(t_i)^{\exp(\zbf_i^{\top} \bgamma)} \nonumber \\
&=& -\exp(\zbf_i^{\top} \bgamma) S_0(t_i)^{\exp(\zbf_i^{\top} \bgamma)-1} (d/dt) S_0(t_i) \nonumber \\
&=& \exp(\zbf_i^{\top} \bgamma) S_0(t_i)^{\exp(\zbf_i^{\top} \bgamma)-1} f_0(t_i) \nonumber \\
&=& \exp(\zbf_i^{\top} \bgamma) S_0(t_i)^{\exp(\zbf_i^{\top} \bgamma)-1} S_0(t_i) h_0(t_i)  \nonumber \\
&=& \exp(\zbf_i^{\top} \bgamma) h_0(t_i) S_0(t_i)^{\exp(\zbf_i^{\top} \bgamma)}.   \nonumber 
\end{eqnarray}

\vspace{-0.3cm}

\noindent It follows that the log-likelihood is:

\vspace{-1cm}

\begin{eqnarray}
\ell(\bxi; \mathcal{D}):&=& \log \mathcal{L}(\bxi; \mathcal{D}) \nonumber \\
&=& \sum_{i=1}^n \tau_i \Big(\log p(\xbf_i)+\zbf_i^{\top} \bgamma + \log h_0(t_i)+\exp(\zbf_i^{\top} \bgamma) \log S_0(t_i) \Big) \nonumber \\
&&+ (1-\tau_i) \log \Big(1-p(\xbf_i)+p(\xbf_i) S_0(t_i)^{\exp(\zbf_i^{\top} \bgamma)} \Big). \nonumber 
\end{eqnarray}

\vspace{-0.2cm}

\noindent Using the B-spline approximations for the baseline quantities, we get:

\vspace{-1cm} 

\begin{eqnarray} 
\ell(\bxi; \mathcal{D})&\approx& \sum_{i=1}^n \tau_i \Bigg(\log p(\xbf_i)+\zbf_i^{\top} \bgamma + \btheta^{\top} b(t_i)-\exp(\zbf_i^{\top} \bgamma) \sum_{j=1}^{j(t_i)} \exp\big(\btheta^{\top} b(s_j)\big)\Delta \Bigg) \nonumber \\
&&+ (1-\tau_i) \log \Bigg(1-p(\xbf_i)+p(\xbf_i) \exp\Big(-\exp(\zbf_i^{\top} \bgamma)\sum_{j=1}^{j(t_i)} \exp\Big(\btheta^{\top} b(s_j)\Big) \Delta \Big) \Bigg). \nonumber 
\end{eqnarray}

\noindent Let us denote by $g_i(\bxi)$ the contribution of the $i$th unit to the log-likelihood:

\vspace{-1cm}

\begin{eqnarray}
g_i(\bxi)&=&\tau_i \Bigg(\log p(\xbf_i)+\zbf_i^{\top} \bgamma + \btheta^{\top} b(t_i)-\exp(\zbf_i^{\top} \bgamma) \sum_{j=1}^{j(t_i)} \exp\big(\btheta^{\top} b(s_j)\big)\Delta \Bigg) \nonumber \\
&&+ (1-\tau_i) \log \Bigg(1-p(\xbf_i)+p(\xbf_i) \exp\Big(-\exp(\zbf_i^{\top} \bgamma)\sum_{j=1}^{j(t_i)} \exp\Big(\btheta^{\top} b(s_j)\Big) \Delta \Big) \Bigg), \nonumber 
\end{eqnarray}

\noindent so that the log-likelihood can be compactly written as $\ell(\bxi; \mathcal{D}) \approx \sum_{i=1}^n g_i(\bxi)$. Using Bayes' theorem, the conditional posterior of the latent vector is (up to a proportionality constant):

\vspace{-1cm} 

\begin{eqnarray} \label{condpost}
p(\bxi \vert \lambda, \mathcal{D}) &\propto& \mathcal{L}(\bxi; \mathcal{D}) p(\bxi \vert \lambda) \nonumber \\
&\propto& \exp\Big(\ell(\bxi; \mathcal{D})-0.5 \bxi^{\top} Q_{\bxi}(\lambda) \bxi \Big) \nonumber \\
&\propto& \exp\left(\sum_{i=1}^n g_i(\bxi)-0.5 \bxi^{\top} Q_{\bxi}(\lambda) \bxi \right). 
\end{eqnarray}

\noindent A second-order Taylor expansion of the log-likelihood yields a quadratic form in the latent vector $\bxi$ and hence can be used to obtain a Laplace approximation to \eqref{condpost} as shown in \hyperref[appendA]{Appendix A}. In what follows, we denote by $\widetilde{p}_G(\bxi \vert \lambda, \mathcal{D})=\mathcal{N}_{\text{dim}(\bxi)}(\bxi^*(\lambda), \Sigma_{\bxi}^*(\lambda))$ the Laplace approximation to $p(\bxi \vert \lambda, \mathcal{D})$ for a given value of $\lambda$.

\subsection{Approximate posterior of the penalty parameter}

\noindent The conditional posterior in \eqref{condpost} is a function of the penalty parameter $\lambda$. In a frequentist setting, an ``optimal'' value for $\lambda$ is generally obtained by means of the Akaike information criterion (AIC) or (generalized) cross-validation. From a Bayesian perspective, $\lambda$ is random and its associated posterior distribution is of crucial importance for optimal smoothing. Mathematically, the posterior of $\lambda$ is given by:

\vspace{-1cm} 

\begin{eqnarray} \label{lambpost1}
p(\lambda \vert \mathcal{D}) \propto \frac{\mathcal{L}(\bxi; \mathcal{D}) p(\bxi \vert \lambda) p(\lambda)}{p(\bxi \vert \lambda, \mathcal{D})}.
\end{eqnarray}

\noindent Using the Laplace approximation to $p(\bxi \vert \lambda, \mathcal{D})$ and replacing the latent vector by its modal value $\bxi^*(\lambda)$ from the Laplace approximation, the marginal posterior in \eqref{lambpost1} is approximated in the spirit of \cite{tierney1986accurate}:

\vspace{-1cm} 

\begin{eqnarray} 
\widetilde{p}(\lambda \vert \mathcal{D}) &\propto& \frac{\exp(\ell(\bxi; \mathcal{D})) p(\bxi \vert \lambda) p(\lambda)}{\widetilde{p}_G(\bxi \vert \lambda, \mathcal{D})}\Big \vert_{\bxi=\bxi^*(\lambda)} \nonumber \\
&\propto&  \sqrt{\det(Q_{\bxi}(\lambda)) \det(\Sigma^*_{\bxi}(\lambda))} \exp\left(\sum_{i=1}^n g_i(\bxi^*(\lambda))-0.5 \bxi^{*T}(\lambda) Q_{\bxi}(\lambda) \bxi^*(\lambda)\right) \lambda^{a_{\lambda}-1} \exp(-b_{\lambda} \lambda). \nonumber 
\end{eqnarray}

\noindent For numerical reasons it is more appropriate to work with the log transformed penalty parameter $v=\log(\lambda)$ as the latter is unbounded. Using the transformation method for random variables, one obtains the following approximated (log) posterior for $v$:

\vspace{-1cm} 

\begin{eqnarray} \label{logpostv} 
\log \widetilde{p}(v \vert \mathcal{D}) &\dot{=}& \sum_{i=1}^n  g_i(\bxi^*(v))-0.5 \bxi^{*T}(v) Q_{\bxi}(v) \bxi^*(v)+0.5 (\log \det(Q_{\bxi}(v))+ \log\det(\Sigma_{\bxi}^*(v))) \nonumber \\
&&+a_{\lambda} v - b_{\lambda} \exp(v),
\end{eqnarray}

\begin{figure}[h!]
\centering
\includegraphics[width=10cm, height=7cm]{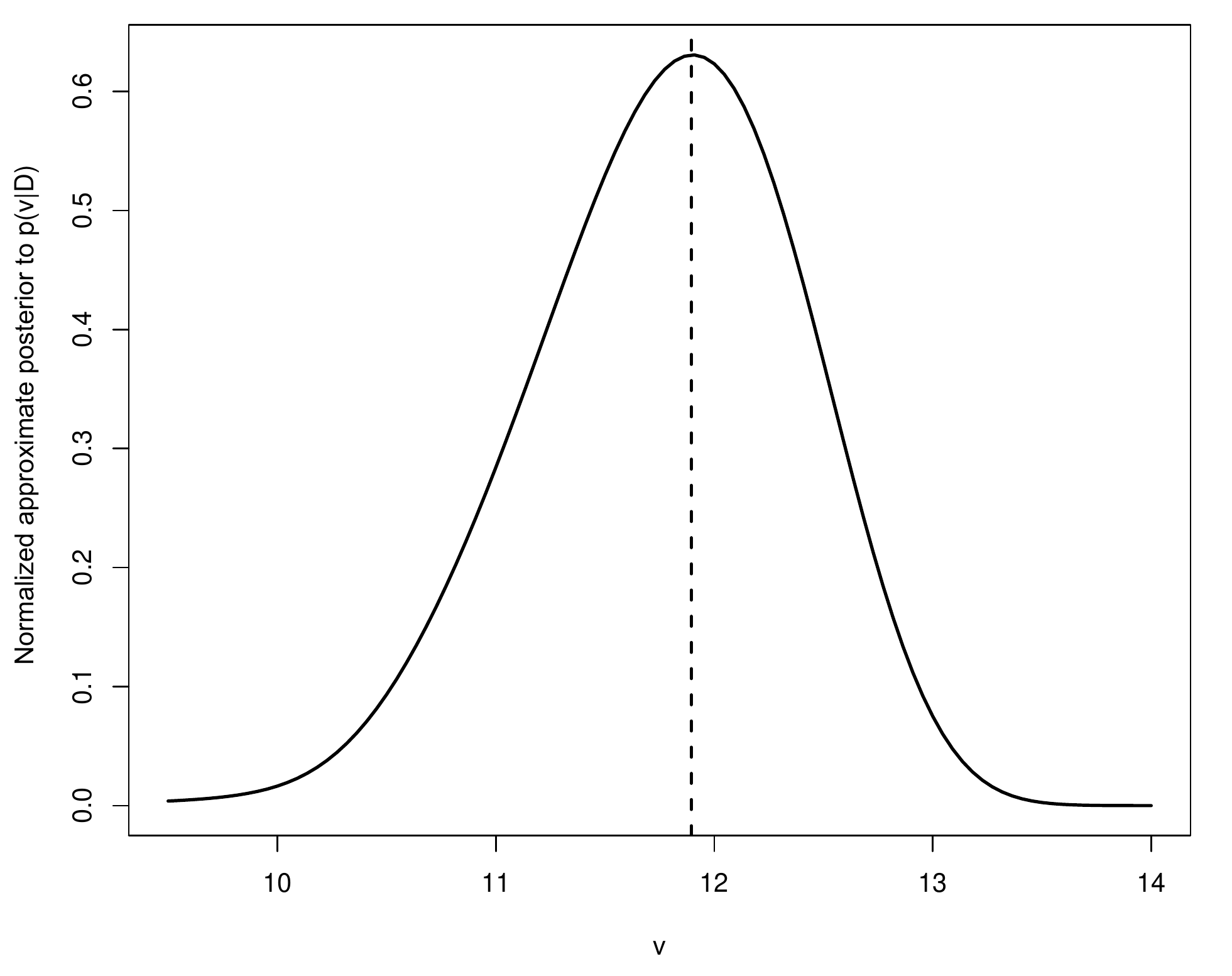}	\caption{Approximated (normalized) posterior of the log penalty $v$. The dashed line is the modal value $v^*$ obtained with the bracketing algorithm using a step size $\delta=0.01$.}
\label{pvmap}
\end{figure}

\vspace{-0.35cm}

\noindent where $\dot{=}$ denotes equality up to an additive constant. Approximation \eqref{logpostv} provides a good starting point for various strategies to explore the posterior penalty space. A possibility is to use grid-based approaches \citep{rue2009approximate,gressani2018fast} or  MCMC algorithms \citep{yoon2011inference,gressani2021laplace} as often encountered in models with a multidimensional penalty space. In latent Gaussian models, the posterior penalty typically satisfies suitable regularity conditions such as unimodality \citep{gomez2017markov} and not a ``too-large'' deviation from Gaussianity. This suggests to use a simple and yet efficient type of bracketing algorithm to compute the (approximate) posterior mode $v^*$ of $\log \widetilde{p}(v \vert \mathcal{D})$. Starting with an arbitrarily ``large'' value, say $v_0=15$, the algorithm moves in the left direction with a fixed step size $\delta$, i.e.\ at the $m$th iteration $v_m=v_{m-1}-\delta$. Movement in the left direction continues until reaching $v_{\widetilde{m}}$, the point at which the target function starts to point downhill $\widetilde{p}(v_{\widetilde{m}}\vert \mathcal{D})<\widetilde{p}(v_{\widetilde{m}}+\delta\vert \mathcal{D})$. The approximated modal value is then $v^*=v_{\widetilde{m}}+\delta/2$. \hyperref[pvmap]{Figure 1} illustrates the normalized approximate posterior to $p(v\vert \mathcal{D})$ from a simulated example along with the modal value (dashed line) obtained with a step size $\delta=0.01$.

\subsection{Approximate credible intervals}

\noindent The Laplace approximation to the conditional posterior of the latent vector evaluated at the (approximated) modal posterior value $v^*$ (cf. Section 2.4) is denoted by $\widetilde{p}_G(\bxi \vert v^*, \mathcal{D})=\mathcal{N}_{\text{dim}(\bxi)}\big(\bxi^*(v^*), \Sigma_{\bxi}^*(v^*)\big)$ and a point estimate for $\bxi$ is taken to be the mean/mode $\bxi^*(v^*)$ with associated variance-covariance matrix $\Sigma_{\bxi}^*(v^*)$. To ensure that the estimated baseline survival function $\widehat{S}_0(\cdot)$ ``lands'' smoothly on the horizontal asymptote at 0 near the end of the follow-up, we constrain the last B-spline coefficient by fixing $\theta_K=1$. A major advantage of LPSMC is that credible intervals for (differentiable functions of) latent variables can be straightforwardly obtained starting from $\widetilde{p}_G(\bxi \vert v^*, \mathcal{D})$.

\noindent \textbf{Credible interval for latent variables}

\noindent A $(1-\alpha) \times 100\%$ (approximate) credible interval for a latent variable $\xi_h \in \bxi$ follows easily from the fact that the Laplace approximated posterior to $\xi_h$ is $\widetilde{p}_G(\xi_h \vert v^*, \mathcal{D})=\mathcal{N}_{1}\big(\xi_h^*, \sigma^{2*}_{\xi_h}\big)$, where $\xi_h^*$ is the $h$th entry of the vector $\bxi^*(v^*)$ and $\sigma^{2*}_{\xi_h}$ is the $h$th entry on the main diagonal of $\Sigma_{\bxi}^*(v^*)$. It follows that a quantile-based $(1-\alpha) \times 100\%$ credible interval for $\xi_h$ is:

\vspace{-1cm}

\begin{eqnarray}
\text{CI}_{\xi_h}=\xi_h^* \pm z_{\alpha/2} \sqrt{\sigma^{2*}_{\xi_h}}, \nonumber 
\end{eqnarray}

\vspace{-0.3cm}

\noindent where $ z_{\alpha/2}$ is the $\alpha/2-$upper quantile of a standard normal variate.

\vspace{0.5cm}

\noindent \textbf{Credible interval for the incidence $p(\xbf)$ and cure rate $1-p(\xbf)$}

\noindent The incidence of the mixture cure model is a function of the latent vector $\breve{\bbeta}$ and hence an appropriate approach to derive credible intervals for $p(\xbf)$ is through using a ``delta'' method. In particular, let us consider the following differentiable function of the probability to be uncured $g(\breve{\bbeta} \vert \xbf)=\log\big(-\log p(\xbf)\big)=\log\big(\log(1+\exp(-\beta_0-\xbf^{\top} \bbeta))\big)$, where $\xbf$ is a known profile of the covariate vector. The Laplace approximated posterior to vector $\breve{\bbeta}$ is known to be $\widetilde{p}_G(\breve{\bbeta} \vert v^*, \mathcal{D})=\mathcal{N}_{\text{dim}(\breve{\bbeta})}\big(\breve{\bbeta}^*, \Sigma^{*}_{\breve{\bbeta}}\big)$ with mean vector $\breve{\bbeta}^*=\breve{\bbeta}^*(v^*)$ and covariance matrix $\Sigma^{*}_{\breve{\bbeta}}=\Sigma^{*}_{\breve{\bbeta}}(v^*)$. The delta method operates via a first-order Taylor expansion of $g(\breve{\bbeta} \vert \xbf)$ around $\breve{\bbeta}^*$:

\vspace{-1.5cm} 

\begin{eqnarray} \label{Taylorbeta}
g(\breve{\bbeta} \vert \xbf) \approx g(\breve{\bbeta}^* \vert \xbf)+(\breve{\bbeta}-\breve{\bbeta}^*)^{\top} \nabla g(\breve{\bbeta} \vert \xbf)\big\vert_{\breve{\bbeta}=\breve{\bbeta}^*},
\end{eqnarray}

\vspace{-0.2cm}

\noindent with gradient:

\vspace{-1.4cm} 

\begin{eqnarray}
\nabla g(\breve{\bbeta} \vert \xbf)&=&\left(\frac{\partial g(\breve{\bbeta} \vert \xbf)}{\partial \beta_0},\dots, \frac{\partial g(\breve{\bbeta} \vert \xbf)}{\partial \beta_p}\right)^{\top} \nonumber \\
&=&\frac{-(1-p(\xbf))}{\log\big(1+\exp(-\beta_0-\xbf^{\top} \bbeta)\big)}\ (1,\xbf^{\top})^{\top}. \nonumber 
\end{eqnarray}

\vspace{-0.2cm}

\noindent Note that $g(\breve{\bbeta} \vert \xbf)$ in \eqref{Taylorbeta} is still Gaussian as it is a linear combination of a random vector $\breve{\bbeta}$ that is a posteriori (approximately) Gaussian due to the Laplace approximation with mean $\mathbb{E}(g(\breve{\bbeta} \vert \xbf))\approx g(\breve{\bbeta}^* \vert \xbf)$ and covariance matrix $\mathbb{V}(g(\breve{\bbeta} \vert \xbf))\approx \nabla^{\top} g(\breve{\bbeta} \vert \xbf)\vert_{\breve{\bbeta}=\breve{\bbeta}^*} \Sigma^*_{\breve{\bbeta}} \nabla g(\breve{\bbeta} \vert \xbf)\vert_{\breve{\bbeta}=\breve{\bbeta}^*}$. This suggests to write the approximated posterior of $g(\breve{\bbeta} \vert \xbf)$ as:

\vspace{-1cm} 

\begin{eqnarray}
\big(g(\breve{\bbeta} \vert \xbf) \vert \mathcal{D}\big) \sim \mathcal{N}_1\left( g(\breve{\bbeta}^* \vert \xbf),  \nabla^{\top} g(\breve{\bbeta} \vert \xbf)\vert_{\breve{\bbeta}=\breve{\bbeta}^*} \Sigma^*_{\breve{\bbeta}} \nabla g(\breve{\bbeta} \vert \xbf)\vert_{\breve{\bbeta}=\breve{\bbeta}^*}\right) \nonumber 
\end{eqnarray}

\vspace{-0.2cm}

\noindent and so an approximate quantile-based $(1-\alpha) \times 100\%$ credible interval for $g(\breve{\bbeta} \vert \xbf)$ is:

\vspace{-1cm} 

\begin{eqnarray} \label{CIp}
CI_{g(\breve{\bbeta} \vert \xbf)}= g(\breve{\bbeta}^* \vert \xbf) \pm z_{\alpha/2} \sqrt{\nabla^{\top} g(\breve{\bbeta} \vert \xbf)\vert_{\breve{\bbeta}=\breve{\bbeta}^*} \Sigma^*_{\breve{\bbeta}} \nabla g(\breve{\bbeta} \vert \xbf)\vert_{\breve{\bbeta}=\breve{\bbeta}^*}}. 
\end{eqnarray}

\vspace{-0.2cm}

\noindent Multiplying the values in the interval \eqref{CIp} by $\exp(-\exp(\cdot))$ gives us the desired credible interval for the incidence $p(\xbf)$. If a credible interval for the cure rate $1-p(\xbf)$ is required, simply use the transformation $g(\breve{\bbeta} \vert \xbf)= \log\big(- \log(1-p(\xbf))\big)=\log\big(\log(1+\exp(\beta_0+\xbf^{\top} \bbeta))\big)$ with gradient $\nabla g(\breve{\bbeta} \vert \xbf)= \Big(p(\xbf)/\log(1+\exp(\beta_0+\xbf^{\top}\bbeta))\Big) (1,\xbf^{\top})^{\top}$.

\vspace{0.5cm}

\noindent \textbf{Credible interval for $S_0(\cdot)$ and $S_u(\cdot \vert \zbf)$}

\noindent Let us denote by $t_q$ the $q$th quantile of the distribution of the survival time $T$ at baseline by $t_q=\inf\{t \vert S_0(t) \leq 1-q\}$. The ``delta'' method can also be used to compute an approximate credible interval for $S_0(\cdot)$ at $t_q$ by using a $\log(-\log(\cdot))$ transformation $g(\btheta \vert t_q)=\log(-\log(S_0(t_q)))=\log\big(\sum_{j=1}^{j(t_q)} \exp(\btheta^{\top} b(s_j)) \Delta \big)$. Starting from the Laplace approximated posterior $\widetilde{p}_G(\btheta^* \vert v^*, \mathcal{D})=\mathcal{N}_{\text{dim}(\btheta)}(\btheta^*, \Sigma_{\btheta}^*)$, one can show that the resulting credible interval for $g(\btheta \vert t_q)$ is:

\vspace{-1cm}

\begin{eqnarray} \label{CIgS0}
CI_{g(\btheta \vert t_q)}= g(\btheta^* \vert t_q) \pm z_{\alpha/2} \sqrt{\nabla^{\top} g(\btheta \vert t_q)\vert_{\btheta=\btheta^*} \Sigma^*_{\btheta} \nabla g(\btheta \vert t_q)\vert_{\btheta=\btheta^*}}, 
\end{eqnarray}

\noindent where $\nabla g(\btheta \vert t_q)\vert_{\btheta=\btheta^*}$ is the gradient of $g(\btheta \vert t_q)$ with respect to $\btheta$ evaluated at $\btheta^*$ and can be found in \cite{gressani2018fast} Appendix C. Applying the inverse transformation $\exp(-\exp(\cdot))$ to \eqref{CIgS0} yields the desired $(1-\alpha)\times 100 \%$ credible interval for $S_0(\cdot)$ at $t_q$.\\
\indent The same approach is used to construct credible intervals for the survival function of the uncured $S_u(t \vert \zbf)=S_0(t)^{\exp(\zbf^{\top} \bgamma)}$ at $t_q=\inf\{t \vert S_u(t \vert \zbf) \leq 1-q\}$ for a given covariate profile $\zbf$. Applying the $\log(-\log(\cdot))$ transform yields $g(\btheta, \bgamma \vert t_q, \zbf)=\zbf^{\top}\bgamma+\log\big(\sum_{j=1}^{j(t_q)} \exp(\btheta^{\top} b(s_j)) \Delta \big)$ with gradient:

\vspace{-0.7cm}

\begin{eqnarray}
\nabla g(\btheta, \bgamma \vert t_q, \zbf)&=&\left(\frac{\partial g(\btheta, \bgamma \vert t_q, \zbf)}{\partial \theta_1},\dots,\frac{\partial g(\btheta, \bgamma \vert t_q, \zbf)}{\partial \theta_K},\frac{\partial g(\btheta, \bgamma \vert t_q, \zbf)}{\partial \gamma_1},\dots,\frac{\partial g(\btheta, \bgamma \vert t_q, \zbf)}{\partial \gamma_q} \right)^{\top} \nonumber \\
\nonumber \\
&=& \begin{pmatrix}
\left(\sum_{j=1}^{j(t_q)} \exp(\btheta^{\top} b(s_j)) \Delta \right)^{-1} \sum_{j=1}^{j(t_q)} \exp(\btheta^{\top} b(s_j)) b_1(s_j) \Delta \\
\vdots \\
\left(\sum_{j=1}^{j(t_q)} \exp(\btheta^{\top} b(s_j)) \Delta \right)^{-1} \sum_{j=1}^{j(t_q)} \exp(\btheta^{\top} b(s_j)) b_K(s_j) \Delta \\
z_1 \\
\vdots \\
z_q
\end{pmatrix}. \nonumber 
\end{eqnarray}

\newpage 

\noindent The resulting credible interval for $g(\btheta, \bgamma \vert t_q, \zbf)$ is:

\vspace{-1cm}

\begin{eqnarray} \label{CIgSu}
CI_{g(\btheta, \bgamma \vert t_q, \zbf)}= g(\btheta^*, \bgamma^* \vert t_q, \zbf) \pm z_{\alpha/2} \sqrt{\nabla^{\top} g(\btheta, \bgamma \vert t_q, \zbf)\vert_{\btheta=\btheta^*, \bgamma=\bgamma^*} \Sigma^*_{\btheta,\bgamma} \nabla g(\btheta, \bgamma \vert t_q, \zbf)\vert_{\btheta=\btheta^*, \bgamma=\bgamma^*}}, 
\end{eqnarray}

\noindent where $\Sigma^*_{\btheta,\bgamma}$ is the covariance matrix of the vector $(\btheta^{\top},\bgamma^{\top})^{\top}$ obtained from the Laplace approximation. Finally, an $\exp(-\exp(\cdot))$ transform on \eqref{CIgSu} gives a $(1-\alpha)\times 100 \%$ credible interval for $S_u(t\vert \zbf)$ at $t_q$ for a given covariate vector $\zbf$.

\section{Simulation study}

\noindent To measure the statistical performance of the LPSMC methodology in a mixture cure model setting, we consider a numerical study where survival data is generated according to different cure and censoring rates. Generation of survival data for the $i$th subject is as follows. The incidence part is generated from a logistic regression function with two covariates $p(\Xbf_i)=1/(1+\exp(-\beta_0-\beta_1 X_{i1}-\beta_2 X_{i2}))$, where $X_{i1}$ is a standard normal variate and $X_{i2} \sim \text{Bern}(0.5)$ is a Bernoulli random variable. The cure status is generated as a Bernoulli random variable with failure probability $p(\Xbf_i)$, i.e.\ $B_i \sim \text{Bern}(p(\Xbf_i))$. Survival times $T_i$ for the uncured subjects ($B_i=1$) are obtained from a Weibull Cox proportional hazards model and are truncated at $\tau_0=8$. The latency is given by $S_u(t_i \vert \Zbf_i)=\exp(-\nu t_i^{\varrho}\exp(\gamma_1 Z_{i1}+\gamma_2 Z_{i2}))$, with scale parameter $\nu=0.25$  and shape $\varrho=1.45$. The covariate vector $\Zbf_i=(Z_{i1},Z_{i2})^{\top}$ is independent of $\Xbf_i$. We assume that $Z_{i1}$ follows a standard Gaussian distribution and $Z_{i2}\sim \text{Bern}(0.4)$. For the cured subject ($B_i=0$), the theoretically infinite survival times are replaced by a large value, say $T_i=20,000$. The censoring time $C_i$ is independent of the vector $(\Xbf_i^{\top}, \Zbf_i^{\top}, T_i)^{\top}$ and is generated from an exponential distribution with density function $f(c_i)=\mu_c \exp(-\mu_c c_i)$ truncated at $\tau_1=11$. We consider samples of sizes $n=300$ and $n=600$ and simulate survival data with two scenarios for the coefficients $\bbeta$, $\bgamma$ and $\mu_c$, yielding different censoring and cure rates. In both scenarios (almost) all the observations in the plateau of the \cite{kaplan1958nonparametric} estimator are cured, such that the simulated data are representative of the practical real case scenarios for which mixture cure models are used. \hyperref[tab1]{Table 1} provides a summary of the two considered scenarios.\\
\indent We specify $15$ cubic B-splines in the interval $[0,t_u]$ with upper bound fixed at $t_u=11$ and a third order penalty to counterbalance model flexibility. In the bracketing algorithm, we use a step size $\delta=0.2$ to compute the (approximate) modal posterior log penalty value $v^*$. For each scenario, we simulate $S=500$ replications and compute the bias, empirical standard error (ESE), root mean square error (RMSE) and coverage probabilities for $90\%$ and $95\%$ credible intervals.

\vspace{0.3cm}

\begin{table}[!h] 
\centering 
\begin{tabular}{c c c c c c c c c c}
\hline 
Scenario & $\beta_0$ & $\beta_1$ & $\beta_2$ & $\gamma_1$ & $\gamma_2$ & Cure & $\mu_c$ & Censoring & Plateau \\
\hline 
1 & 0.70 & -1.15 & 0.95 & -0.10 & 0.25 & $28.8\%$ & 0.16 & $48.5\%$ & $9.6\%$ \\
2 & 1.25 & -0.75 & 0.45 & -0.10 & 0.20 & $21.0\%$ & 0.05 & $29.3\%$ & $14.4\%$ \\
\hline 
\end{tabular}
\caption{Parameters for the incidence, latency and censoring rate yielding different cure and censoring levels. The last column indicates the percentage of observations in the plateau.}
\label{tab1}
\end{table} 

\vspace{-0.3cm}

\begin{figure}[h!]
\centering
\includegraphics[width=16cm, height=11.6cm]{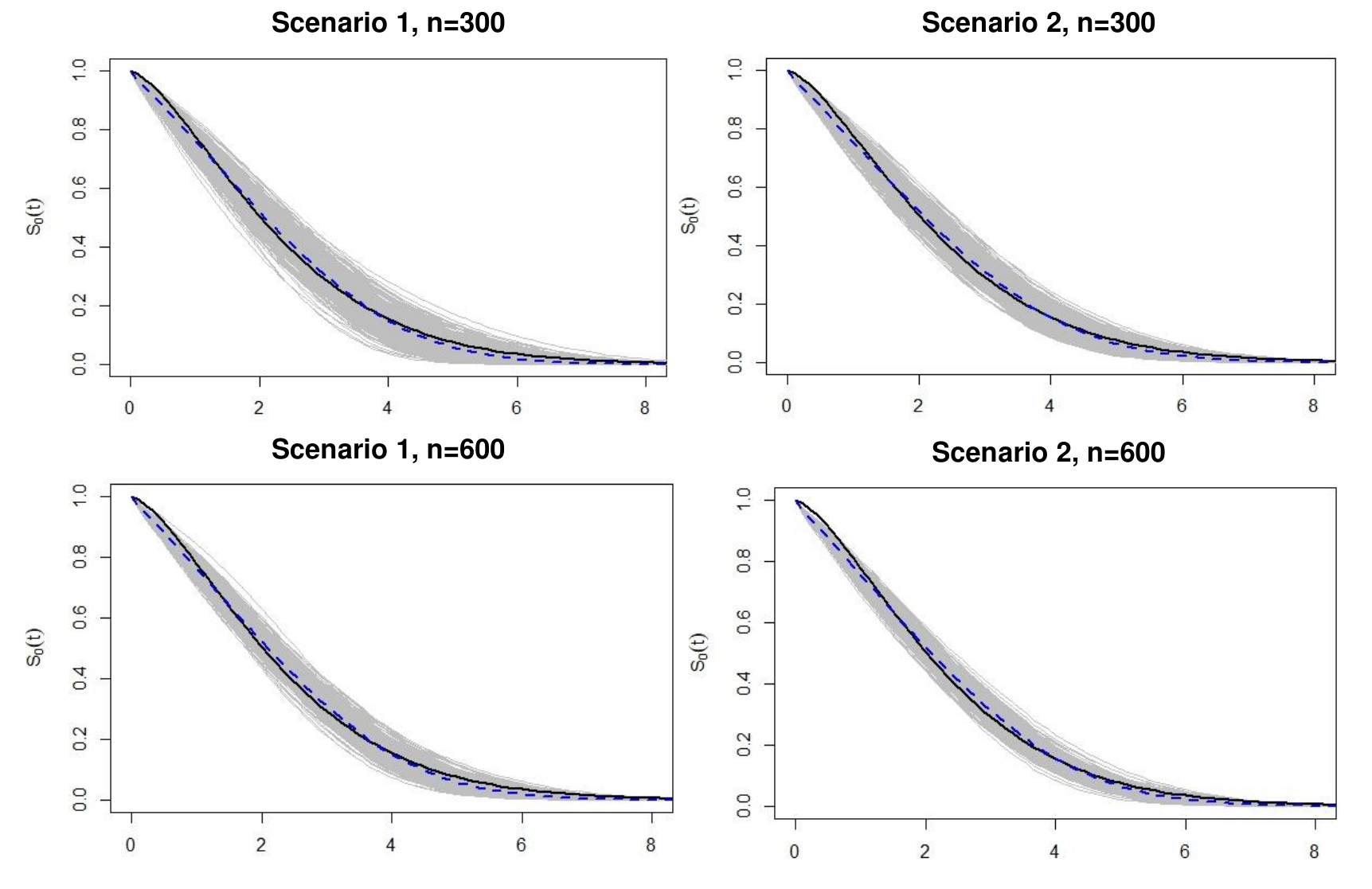}	\caption{Estimated baseline survival curves (gray) for $S=500$ replications under different scenarios. The black curve is the target baseline and the dashed curve the pointwise median of the 500 gray curves.}
\label{fig2}
\end{figure}

\begin{figure}[h!]
	\centering
	\includegraphics[width=9cm, height=6.7cm]{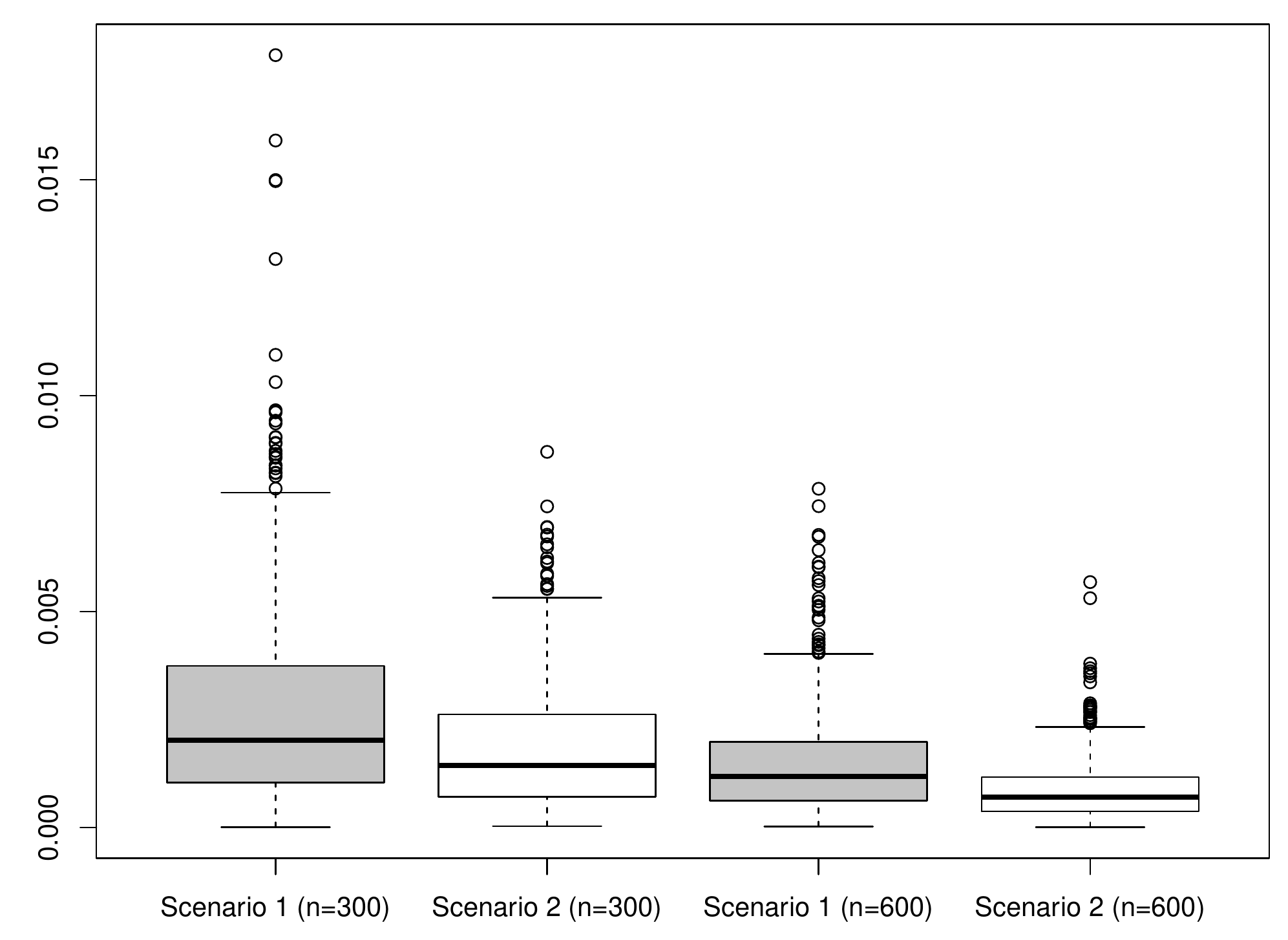}	\caption{Boxplots of the ASE for the incidence in different scenarios.}
	\label{fig3}
\end{figure}

\noindent Results are summarized in \hyperref[tab2]{Table 2}. The bias is negligible and the ESE and RMSE decrease with larger sample size, as expected. In addition, the estimated coverage probabilities are close to their respective nominal value in all scenarios. The LPSMC methodology is extremely fast as it takes $\sim$ 0.7 seconds to fit a model with an algorithm coded in $\bf{R}$ with an Intel Xeon E-2186M processor at 2.90GHz. \hyperref[fig2]{Figure 2} shows the estimated baseline survival curves (gray), the target (solid) and the pointwise median of the $500$ curves (dashed) for the different scenarios.\\
\indent Another performance measure for the incidence of the model is obtained by computing the Average Squared Error (ASE) of $\widehat{p}(\xbf)$ defined as $ASE(\widehat{p})=M^{-1} \sum_{m=1}^M \big(\widehat{p}(\xbf)-p(\xbf)\big)^2$. The latter quantity is computed on triplets of covariate values $\boldsymbol{x}_m=(1,x_{m1},x_{m2})^{\top}$ for $m=1,\dots,M$, where the set of couples $\{(x_{m1}, x_{m2})\}_{m=1}^M$ equals the Cartesian product between an equidistant grid in $[-1.50, 1.50]$ with step size $0.001$ (for variable $x_1$) and the set $\{0,1\}$ (for $x_2$). Boxplots for the ASE in the different scenarios are displayed in \hyperref[fig3]{Figure 3}. Coverage probability of the $90\%$ and $95\%$ (approximate) credible interval for the incidence $p(\xbf)$ (and cure rate $1-p(\xbf)$) at the mean covariate profile $\bar{\xbf}=(1,0,0.5)^{\top}$ as computed from \eqref{CIp} have also been measured and are close to their nominal value in all scenarios.

\newpage 

\vspace{0.5cm}

\newpage 

\newcommand{\minsp}{\hspace{-0.15cm}}
\begin{table}[!h] 
\centering 
\setlength{\tabcolsep}{12pt}
\begin{tabular}{c c c c c c c c}
\hline
Scenario & Parameters & Mean & Bias & ESE & RMSE & $\text{CP}_{90\%}$ &	$\text{CP}_{95\%}$ \\ 
\hline 
\rowcolor{lightgray} 
\hline 
\cellcolor{white}   & $\beta_0 = \phantom{-}0.70$  & 0.720 & 0.020 & 0.249 & 0.249 & 91.0 & 97.0 \\
& $\beta_1 = -1.15$  & \minsp -1.180 & \minsp -0.030 & 0.240 & 0.242 & 91.6 & 95.0 \\
\rowcolor{lightgray} 
\cellcolor{white} Scenario 1 & $\beta_2 = \phantom{-}0.95$  & 0.953 & 0.003 & 0.390 & 0.390 & 90.6 & 94.2 \\
($n=300$) & $\gamma_1 = -0.10$ & \minsp -0.101 & \minsp -0.001 & 0.092 & 0.092 & 89.8 & 94.4 \\
\rowcolor{lightgray}
\cellcolor{white} & $\gamma_2 = \phantom{-}0.25$ & 0.247 & \minsp -0.003 & 0.185 & 0.184 & 89.0 & 96.2 \\
\hline 
& $\beta_0 = \phantom{-}1.25$  & 1.277 & 0.027 & 0.228 & 0.229 & 91.4 & 95.8 \\
\rowcolor{lightgray}
\cellcolor{white} & $\beta_1 = -0.75$  & \minsp -0.763 & \minsp -0.013 & 0.182 & 0.182 & 91.0 & 95.4 \\
Scenario 2 & $\beta_2 = \phantom{-}0.45$  & 0.429 & \minsp -0.021 & 0.329 & 0.329 & 89.0 & 95.0 \\
\rowcolor{lightgray}
\cellcolor{white} ($n=300$) & $\gamma_1 = -0.10$ & \minsp -0.103 & \minsp -0.003 & 0.074 & 0.074 & 89.2 & 94.4 \\
& $\gamma_2 = \phantom{-}0.20$ & 0.197 & \minsp -0.003 & 0.151 & 0.150 & 87.4 & 95.0 \\
\hline 
\rowcolor{lightgray}
\hline 
\cellcolor{white}   & $\beta_0 = \phantom{-}0.70$  & 0.699 & \minsp -0.001 & 0.184 & 0.184 & 90.0 & 95.6 \\
& $\beta_1 = -1.15$  & \minsp -1.150 &  0.000 & 0.166 & 0.166 & 90.4 & 95.2 \\
\rowcolor{lightgray} 
\cellcolor{white} Scenario 1 & $\beta_2 = \phantom{-}0.95$  & 0.948 & \minsp -0.002 & 0.268 & 0.267 & 90.8 & 95.0 \\
($n=600$) & $\gamma_1 = -0.10$ & \minsp -0.102 & \minsp -0.002 & 0.064 & 0.064 & 89.6 & 95.0 \\
\rowcolor{lightgray}
\cellcolor{white} & $\gamma_2 = \phantom{-}0.25$ & 0.256 & 0.006 & 0.127 & 0.127 & 90.8 & 96.0 \\
\hline
 & $\beta_0 = \phantom{-}1.25$  & 1.241 & \minsp -0.009 & 0.160 & 0.160 & 91.2 & 95.8 \\
\rowcolor{lightgray}
\cellcolor{white} & $\beta_1 = -0.75$  & \minsp -0.744 & 0.006 & 0.129 & 0.129 & 90.0 & 95.4 \\
Scenario 2 & $\beta_2 = \phantom{-}0.45$  & 0.457 &  0.007 & 0.222 & 0.222 & 91.4 & 95.8 \\
\rowcolor{lightgray}
\cellcolor{white} ($n=600$) & $\gamma_1 = -0.10$ & \minsp -0.100 & 0.000 & 0.054 & 0.054 & 88.6 & 95.6 \\
& $\gamma_2 = \phantom{-}0.20$ & 0.200 & 0.000 & 0.105 & 0.105 & 90.6 & 95.2 \\
\hline 
\end{tabular}
\caption{Numerical results for $S=500$ replications of sample size $n=300$ and $n=600$ under two different cure-censoring scenarios.}
\label{tab2}
\end{table} 

\noindent In \hyperref[tab3]{Table 3}, the performance of approximate credible intervals for the baseline survival and survival of the uncured for a mean covariate profile $\bar{\zbf}=(0,0.4)^{\top}$ at selected quantiles of $T$ is shown. Results are for $S=500$ replications of sample size $n=300$ with $K=30$ B-splines. 

\vspace{0.4cm}

\begin{table}[!h] 
\centering 
\setlength{\tabcolsep}{6pt}
\begin{tabular}{c c c c c c c c c c c c}
\hline
& Nominal & Scenario & $t_{0.20}$ & $t_{0.30}$  & $t_{0.40}$ & $t_{0.50}$ & $t_{0.60}$ & $t_{0.70}$ & $t_{0.80}$ & $t_{0.90}$ & $t_{0.95}$ \\
\hline 
&  90 & 1 & 83.2  & 92.4 & 91.4 & 90.6 & 89.8 & 89.8 & 90.6 & 86.0 & 79.6 \\
Baseline &  95 & 1 & 91.0 & 95.6 & 95.0 & 94.2 & 93.6 & 93.8 & 94.4 & 93.2 & 88.2 \\
$S_0(\cdot)$ &  90 & 2 & 76.0 & 91.2 & 93.0 & 93.4 & 90.6 & 91.0 & 91.2 & 84.4  & 80.0 \\
&  95 & 2 & 84.0 & 95.6 & 96.6 & 96.6 & 95.8 & 95.8 & 95.0 & 90.6 & 86.6 \\
\hline 
&  90 & 1 & 81.6 & 93.6 & 93.2 & 89.8 & 89.0 & 88.8 & 89.0 & 90.4 & 83.4 \\
Uncured &  95 & 1 & 92.2 & 96.4 & 97.0 & 95.2 & 94.2 & 94.6 & 94.0 & 95.4 & 90.0 \\
$S_u(\cdot \vert \bar{\zbf})$ &  90 & 2 & 77.4 & 91.8 & 94.2 & 93.4 & 92.0 & 90.6 & 89.8 & 86.0 & 79.4 \\
&  95 & 2 & 85.8 & 95.8 & 97.6 & 97.0 & 95.8 & 95.4 & 94.4 & 92.6 & 86.0 \\
\hline 
\end{tabular}
\caption{Estimated coverage probability of $90\%$ and $95\%$ credible intervals for $S_0(\cdot)$ and $S_u(\cdot \vert \bar{\zbf}=(0,0.4)^{\top})$ 
\label{tab3} with $S=500$ replications of sample size $n=300$ and $K=30$ B-splines.}
\end{table}

\newpage 

\section{Real data applications}

\subsection{ECOG e1684 clinical trial}

\begin{figure}[h!]
\centering
\includegraphics[width=10cm, height=7cm]{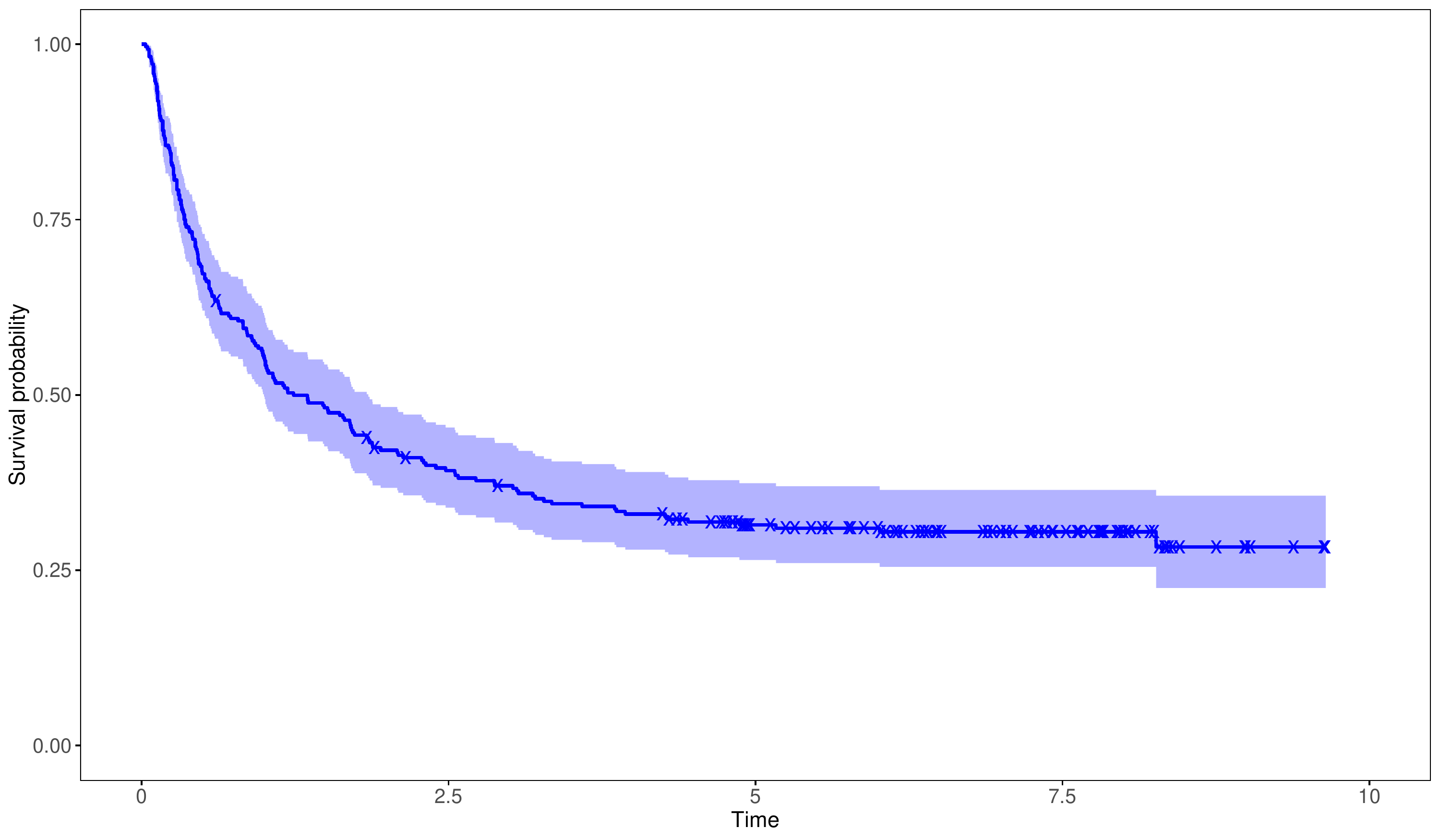}	\caption{Kaplan-Meier estimated survival for the e1684 ECOG dataset. A cross indicates a censored observation.}
\label{fig4}
\end{figure}

\noindent We start by applying the LPSMC methodology to a well-known dataset in the cure literature and compare the estimates with the ones obtained using the benchmark {\bf{smcure}} package \citep{cai2012smcure}. The data comes from the Eastern Cooperative Oncology Group (ECOG) phase III two-arm clinical trial with $n=284$ observations after removing missing data. The aim of the study was to assess whether Interferon alpha-2b (IFN) had a significant effect on relapse-free survival. There were 144 patients receiving the treatment and the remaining patients (140) belonged to the control group. The response of interest is relapse-free survival time measured in years. The following covariates enter both the incidence and latency part of the model. \textit{TRT} is the binary variable indicating whether a patient received the IFN treatment ($TRT=1$) or not ($TRT=0$). Variable \textit{SEX} indicates if a patient is female ($SEX=1$) or male ($SEX=0$). In total the study involves 113 women and 171 men. Finally, \textit{AGE} is a continuous covariate (centered to the mean) indicating the age of patients. \hyperref[fig4]{Figure 4}, shows the Kaplan-Meier estimated survival curve for the e1684 dataset. A plateau is clearly visible indicating the potential presence of a cure fraction, so that a mixture cure model is appropriate to fit this type of data.

\newpage 

\noindent \hyperref[tab4]{Table 4} summarizes the estimation results for the e1684 dataset with LPSMC and {\bf{smcure}}. The estimated coefficients in the two parts of the model (incidence and latency) are of similar magnitude for both approaches. However, the computational times to fit the mixture cure model  drastically differ depending on the method. 

\vspace{0.3cm} 

\begin{table}[!h]
\label{tab4} 
\centering 
\setlength{\tabcolsep}{16pt}
\begin{tabular}{c c c c c}
\hline 
\noalign{\vskip 0.1cm}
Model  & Parameter & Estimate & Sd & CI $90\%$ \\
\hline 
\noalign{\vskip 0.1cm}
LPSMC & \hspace{0.95cm} $\beta_0$ (Intercept) & 1.235 & 0.255 & [\phantom{-}0.817; \phantom{-}1.654]  \\
(Incidence) & \hspace{0.05cm} $\beta_1$ (SEX) & \minsp -0.064 & 0.291 & [-0.542; \phantom{-}0.415] \\
& \hspace{0.05cm} $\beta_2$ (TRT) & \minsp -0.572 & 0.289 & [-1.046; -0.097]\\
& \hspace{0.05cm} $\beta_3$ (AGE) & 0.016 & 0.011 & [-0.003; \phantom{-}0.035]\\
smcure &  \hspace{0.75cm} $\beta_0$ (Intercept) & 1.365 & 0.329 & [\phantom{-}0.823; \phantom{-}1.906]  \\
(Incidence) & $\beta_1$ (SEX) & \minsp -0.087 & 0.333 & [-0.634; \phantom{-}0.461]\\
& $\beta_2$ (TRT)& \minsp -0.588 & 0.343 & [-1.152; -0.025] \\
& $\beta_3$ (AGE) & 0.020 & 0.016 & [-0.006; \phantom{-}0.047]\\
\hline 
\noalign{\vskip 0.1cm}
LPSMC & $\gamma_1$ (SEX) & 0.096 & 0.177 & [-0.195; \phantom{-}0.388] \\
(Latency)&  $\gamma_2$ (TRT) & \minsp -0.131 & 0.179 & [-0.425; \phantom{-}0.163]  \\
&  $\gamma_2$ (AGE) & \minsp -0.007 & 0.006 & [-0.017; \phantom{-}0.003] \\
smcure & $\gamma_1$ (SEX) & 0.099 & 0.175 & [-0.189; \phantom{-}0.388]\\
(Latency)& $\gamma_2$ (TRT) & \minsp -0.154 & 0.177 & [-0.444; \phantom{-}0.137] \\
& $\gamma_3$ (AGE) & \minsp -0.008 & 0.007 & [-0.019; \phantom{-}0.004]\\
\hline 
\end{tabular}
\caption{Estimation results for the e1684 dataset with LPSMC and {\bf{smcure}}. The first column indicates the model part, the second and third columns the parameter and its estimate. The fourth column is the (posterior) standard deviation of the estimate and the last column the $90\%$ confidence/credible interval.}
\end{table}

\noindent It takes approximately 128 seconds to fit the model with {\bf{smcure}}, while LPSMC takes only 1.6 seconds. This speedup factor of $\approx 80$ is mainly due to the bootstrap samples that {\bf{smcure}} uses to compute estimates of the variance of estimated parameters. The LPSMC approach is completely sampling-free and intrinsically accounts for fast computation of credible intervals for the latent variables. Results for both {\bf{smcure}} and LPSMC show that treatment (TRT) has a significant and positive impact on the cure probability (incidence), while IFN treatment has no significative impact in the latency part. In other words, taking \textit{AGE} and \textit{SEX} into account, receiving the IFN treatment decreases the probability to be susceptible but will not postpone relapse among the uncured. Similar findings are reported in \cite{corbiere2007sas} and \cite{legrand2019cure}. 

\newpage 

\subsection{Breast cancer data}

\begin{figure}[h!]
\centering
\includegraphics[width=11cm, height=6.8cm]{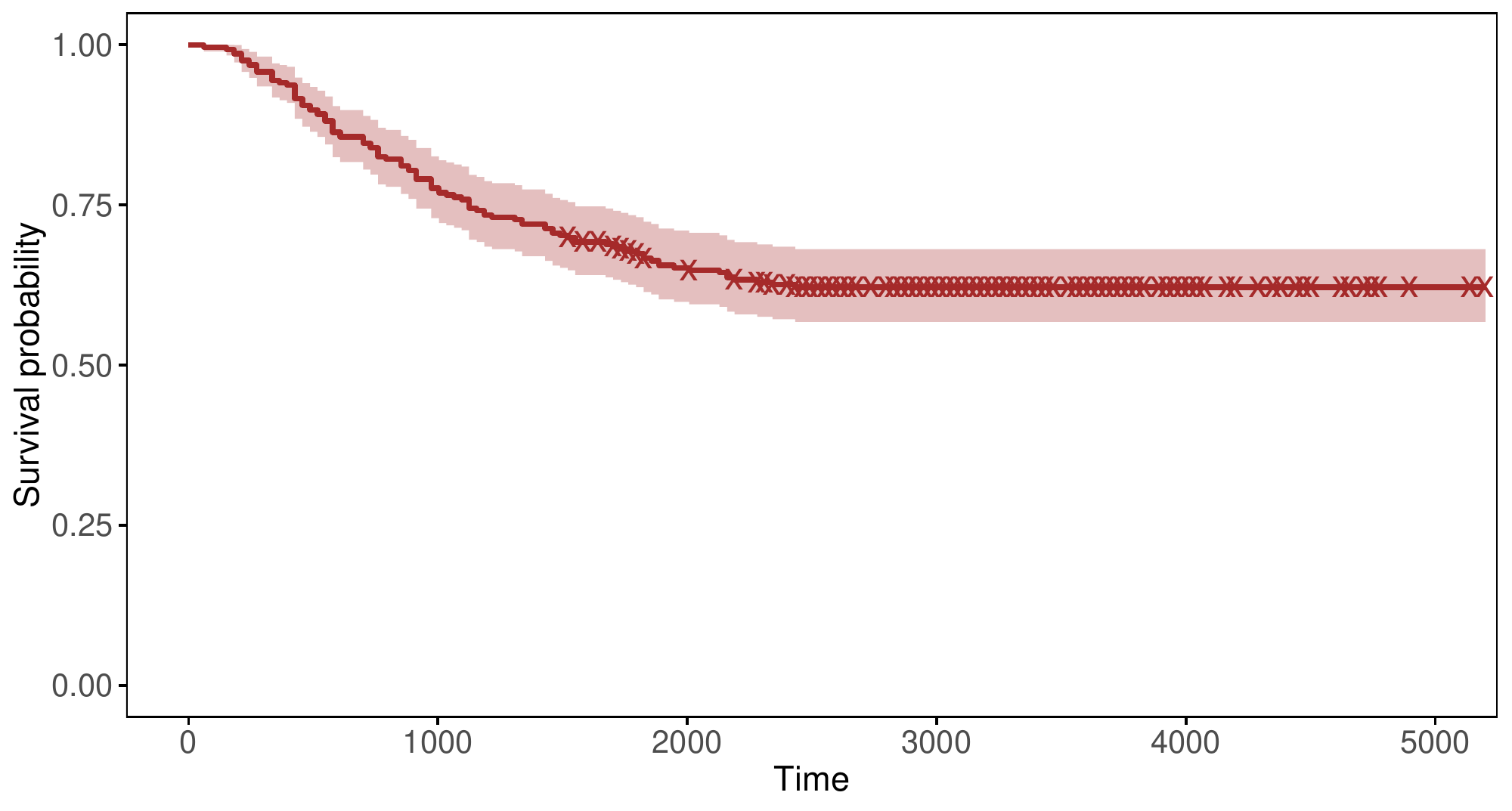}	\caption{Kaplan-Meier estimated survival for the breast cancer data. A cross indicates a censored observation.}
\label{fig5}
\end{figure}

\noindent The second dataset used to illustrate the LPSMC methodology concerns $n=286$ patients with breast cancer studied in \cite{wang2005gene}. Survival time (in days) is defined as the distant-metastasis-free survival (DMFS), i.e.\ time to occurrence of distant metastases or death (whatever happens first) and the event of interest is distant-metastasis. The data is obtained from the {\bf{breastCancerVDX}} package \citep{breastcancer} on Bioconductor (\url{https://bioconductor.org/}). We consider two covariates entering simultaneously in the incidence and latency part, namely the age of patients, ranging between 26 and 83 years with a median of 52 years and the categorical variable Estrogen Receptor \textit{(ER)} (with $ER=0$: if $\leq10$ fmol per mg protein [77 patients] and $ER=1$ otherwise [209 patients]). We use $K=20$ cubic B-splines in $[0,t_u]$, with $t_u=5201$ days, the largest follow-up time. The Kaplan-Meier estimate of the data given in \hyperref[fig5]{Figure 5} emphasizes the existence of a plateau and motivates the use of a mixture cure model. 

\vspace{0.5cm}

\vspace{0.2cm}

\noindent The estimated latent variables with LPSMC are reported in \hyperref[tab5]{Table5}. We see that \textit{AGE} and \textit{ER} have no significant effect on the probability of being uncured (incidence) at significance level $0.05$. However, the latter two covariates significantly affect the survival of the uncured. 

\newpage 

\begin{table}[!h]
\label{tab5} 
\centering 
\setlength{\tabcolsep}{16pt}
\begin{tabular}{c c c c c}
\hline 
\noalign{\vskip 0.1cm}
Model  & Parameter & Estimate & Sd & CI $95\%$ \\
\hline 
\noalign{\vskip 0.1cm}
& $\beta_0$ (Intercept) & \minsp -0.015 & 0.576 & [-1.143; \phantom{-}1.112]  \\
(Incidence) & \hspace{-0.8cm} $\beta_1$ (AGE) & \minsp -0.012 & 0.010 & [-0.031; \phantom{-}0.008] \\
& \hspace{-1cm}$\beta_2$ (ER) & 0.181 & 0.281 & [-0.369; \phantom{-}0.731]\\ 
\hline 
\noalign{\vskip 0.1cm}
(Latency)& \hspace{-0.8cm} $\gamma_1$ (AGE) & \minsp -0.018 & 0.008 & [-0.034; -0.002]  \\
&  \hspace{-1.1cm} $\gamma_2$ (ER) & \minsp -1.063 & 0.243 & [-1.539; -0.586] \\
\hline 
\end{tabular}
\caption{Estimation results for the breast cancer data with LPSMC. The first column indicates the model part, the second and third columns the parameter and its posterior (mean) estimate. The fourth column is the posterior standard deviation of the estimate and the last column the $95\%$ credible interval.}
\end{table}

\begin{figure}[h!]
\centering
\includegraphics[width=11cm, height=6.8cm]{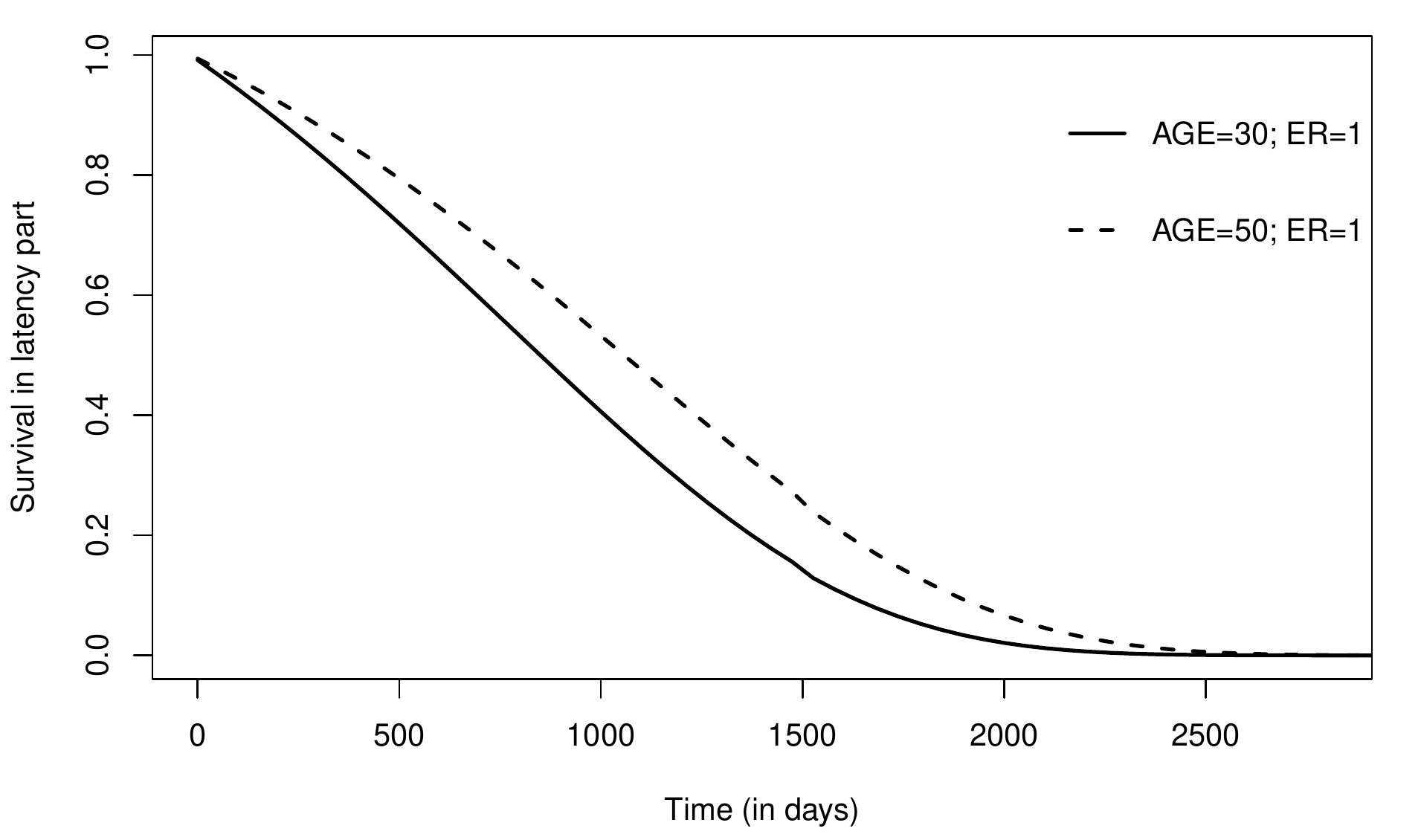}	\caption{Estimated survival of the susceptibles for two age categories and $ER=1$.}
\label{fig6}
\end{figure}

\noindent For instance, a subject that is susceptible to experience metastasis with $ER=1$ has a risk of experiencing the event that is $1/\exp(-1.063)=2.90$ times smaller than the risk with $ER=0$. In \hyperref[fig6]{Figure 6} the estimated survival for the susceptible subjects is shown for two age categories ($30$ years and $50$ years) with $ER=1$. 

\vspace{0.6cm} 

\subsection{ZNA COVID-19 data}

\begin{figure}[h!]
\centering
\includegraphics[width=11cm, height=6.8cm]{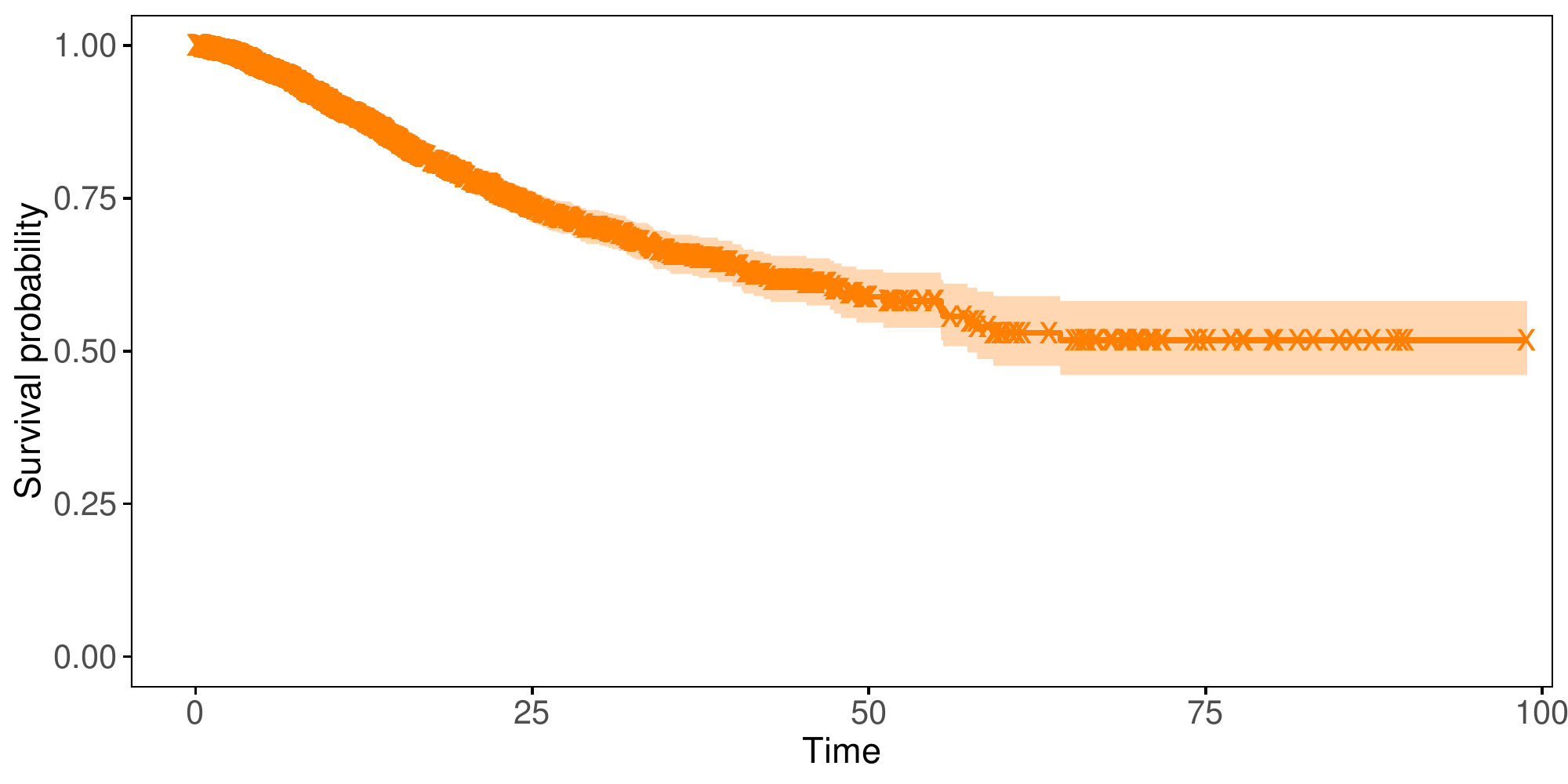}
\caption*{Figure 7.\ Kaplan-Meier estimated survival for the ZNA data. A cross indicates a censored observation.}
\label{fig7}
\end{figure}

\noindent In a third application, we use LPSMC to investigate the impact of age on survival of Covid-19 patients. The data comes from the Ziekenhuis Netwerk Antwerpen (ZNA, Belgium), a network of hospitals in the province of Antwerp. The considered dataset has $n=3258$ patients entering hospitals between March 2020 and April 2021. The follow-up time is defined as the total number of days spent in hospital and receiving COVID-19 care. The outcome of interest is in-hospital death due to Covid-19 as indicated by a binary variable (1 = Dead; 0 = Alive). Among the 3258 patients, 461 (14.15$\%$) experienced in-hospital death and the remaining 2797 (85.85$\%$) are censored due to causes unrelated to the outcome of interest (uninformative censoring). The Kaplan-Meier curve is given in \hyperref[fig7]{Figure 7} and highlights a wide plateau around 0.52, suggesting the existence of a cure fraction and hence motivating a cure model analysis with LPSMC. The age of patients is taken as a covariate both in the incidence and latency part of the model. The mean age is 66 years and the youngest, respectively oldest patient is 1 and 103 years old. We use $K=15$ cubic B-splines between $0$ and the largest follow-up time ($t_u=98.88$) along with a third order penalty.

\vspace{0.5cm}

\vspace{0.1cm}

\noindent \hyperref[tab6]{Table 6} summarizes the estimation results for the model parameters with the LPSMC methodology. We see that \textit{AGE} has a positive and significant effect in the incidence part of the model. In other words, older patients have a larger probability of being uncured and hence a smaller probability to be cured from COVID-19. The estimated cure proportion for different \textit{AGE} categories is shown in \hyperref[tab7]{Table 7}. In the latency part, \textit{AGE} is positive and not significant at the 0.05 level and thus there is no significant effect of \textit{AGE} on the survival of uncured patients.

\vspace{0.2cm}

\begin{table}[!h]
\label{tab6} 
\centering 
\setlength{\tabcolsep}{16pt}
\begin{tabular}{c c c c c}
\hline 
\noalign{\vskip 0.1cm}
Model  & Parameter & Estimate & Sd & CI $95\%$ \\
\hline 
\noalign{\vskip 0.1cm}
& $\beta_0$ (Intercept) & \minsp -2.625 & 0.763 & [-4.120; -1.130]  \\
\noalign{\vskip 0.1cm}
(Incidence) & \hspace{-0.8cm} $\beta_1$ (AGE) & 0.031 & 0.010 & [\phantom{-}0.012; \phantom{-}0.051] \\
\noalign{\vskip 0.1cm}
(Latency)& \hspace{-0.8cm} $\gamma_1$ (AGE) & 0.010 & 0.007 & [-0.004; \phantom{-}0.024]  \\
\hline 
\end{tabular}
\caption{Estimation results for the ZNA data with LPSMC. The first column indicates the model part, the second and third columns the parameter and its posterior (mean) estimate. The fourth column is the posterior standard deviation of the estimate and the last column the $95\%$ credible interval.}
\end{table}

\vspace{0.5cm}

\begin{table}[!h]
\label{tab7} 
\centering 
\setlength{\tabcolsep}{16pt}
\begin{tabular}{c c c}
\hline 
\noalign{\vskip 0.1cm}
\textit{AGE}  & $1-\widehat{p}(\xbf)$ &  CI $95\%$ \\
\hline 
\noalign{\vskip 0.1cm}
20 & 0.880 & [0.695; \phantom{-}0.956]  \\
\noalign{\vskip 0.1cm}
30 & 0.843 &  [0.669; \phantom{-}0.930]  \\
\noalign{\vskip 0.1cm}
50 & 0.741 &  [0.612; \phantom{-}0.833]  \\
\noalign{\vskip 0.1cm}
80 & 0.527 &  [0.463; \phantom{-}0.586]  \\
\hline 
\end{tabular}
\caption{Estimation of the cure proportion for different age categories and associated $95\%$ approximate credible interval.}
\end{table}

\section{Conclusion}

\noindent Approximate Bayesian inference methods are an interesting alternative to classic MCMC algorithms, especially when the latter require long computation times, as can be the case for models with complex likelihoods. In survival analysis, the mixture cure model is an interesting class of models that allows for the existence of a cure fraction and hence goes beyond classic proportional hazards model for which the feature of long-term survivors is absent. In this article, we propose a new approach for fast Bayesian inference in the standard mixture cure model by combining the strength of Laplace approximations to selected posterior distributions of latent variables and P-splines for flexible modeling of baseline smooth quantities. The attractiveness of the LPSMC approach lies in its completely sampling-free framework, with an analytically available gradient and Hessian of the log-likelihood that makes the approach extremely fast from a computational viewpoint. This computational advantage and the relatively straightforward possibility to derive (approximate) credible intervals for functions of latent variables makes it a promising tool for fast analysis of survival data with a cure fraction. A possible interesting extension for future research would be to generalize the LPSMC approach at the level of the incidence with alternative specifications to the standard logistic regression model. For instance, LPSMC could be extended to the single-index/Cox mixture cure model of \cite{amico2019single}. Finally, Laplacian-P-splines can also be extended in cure models with frailties or in the context of competing risks survival data with a cure fraction.

\section*{Software}

R code for the simulation scenarios in Section 3 and the ECOG and breast cancer data applications in Section 4 are publicly available on \url{https://github.com/oswaldogressani/LPSMC}. 

\section*{Conflict of interest}

\noindent The authors declare no conflicts of interest.

\section*{Acknowledgments}

\noindent This project is funded by the European Union's Research and Innovation Action under the H2020 work programme, EpiPose (grand number 101003688). The authors wish to thank the Ziekenhuis Netwerk Antwerpen for granting access to the Covid-19 hospitalization data.

\newpage 

\section*{Appendix A}
\phantomsection{}
\label{appendA}

\newcommand{\gradG}{\nabla g_i(\bxi) \vert_{\bxi=\bxi^{(0)}}}
\newcommand{\hessG}{\nabla^2 g_i(\bxi) \vert_{\bxi=\bxi^{(0)}}}

\noindent Recall that the contribution of the $i$th unit to the log-likelihood is given by:

\vspace{-1cm} 

\begin{eqnarray}
g_i(\bxi)&=&\tau_i \Bigg(\log p(\xbf_i)+\zbf_i^{\top} \bgamma + \btheta^{\top} b(t_i)-\exp(\zbf_i^{\top} \bgamma) \sum_{j=1}^{j(t_i)} \exp\big(\btheta^{\top} b(s_j)\big)\Delta \Bigg) \nonumber \\
&&+ (1-\tau_i) \log \Bigg(1-p(\xbf_i)+p(\xbf_i) \exp\Big(-\exp(\zbf_i^{\top} \bgamma)\sum_{j=1}^{j(t_i)} \exp\Big(\btheta^{\top} b(s_j)\Big) \Delta \Big) \Bigg), \nonumber 
\end{eqnarray}

\vspace{-0.2cm} 

\noindent and using the definition of the population survival function, one has:

\vspace{-1cm}

\begin{eqnarray}
g_i(\bxi)&=&\tau_i \Bigg(\log p(\xbf_i)+\zbf_i^{\top} \bgamma + \btheta^{\top} b(t_i)-\exp(\zbf_i^{\top} \bgamma) \sum_{j=1}^{j(t_i)} \exp\big(\btheta^{\top} b(s_j)\big)\Delta \Bigg) \nonumber \\
&&+ (1-\tau_i) \log \big(S_p(t_i\vert \xbf_i, \zbf_i) \big), \nonumber 
\end{eqnarray}

\vspace{-0.2cm} 

\noindent A second-order Taylor expansion of $g_i(\bxi)$ around and arbitrary point $\bxi^{(0)}$ is given by:

\vspace{-1cm} 

\begin{eqnarray} \label{2Taylor}
g_i(\bxi) &\approx& g_i(\bxi^{(0)})+(\bxi-\bxi^{(0)})^{\top} \gradG + \frac{1}{2} (\bxi-\bxi^{(0)})^{\top} \hessG (\bxi-\bxi^{(0)}) \nonumber \\
&\approx& c+\bxi^{\top} \left(\gradG-\hessG \bxi^{(0)}\right)+\frac{1}{2} \bxi^{\top} \hessG \bxi, 
\end{eqnarray}

\noindent where $c$ is a constant that does not depend on $\bxi$. The gradient $\gradG$ and Hessian $\hessG$ are given by:

\begin{eqnarray}
\nabla g_i(\boldsymbol{\xi})\vert_{\boldsymbol{\xi}=\boldsymbol{\xi}^{(0)}}=
\begin{bmatrix}
\frac{\partial}{\partial \theta_1} g_i(\boldsymbol{\xi}) \\
\vdots\\
\frac{\partial}{\partial \theta_K} g_i(\boldsymbol{\xi})\\
\frac{\partial}{\partial \beta_0} g_i(\boldsymbol{\xi})\\
\vdots \\
\frac{\partial}{\partial \beta_p} g_i(\boldsymbol{\xi})\\
\frac{\partial}{\partial \gamma_1} g_i(\boldsymbol{\xi})\\
\vdots\\
\frac{\partial}{\partial \gamma_q} g_i(\boldsymbol{\xi})\\
\end{bmatrix}_{\boldsymbol{\xi}=\boldsymbol{\xi}^{(0)}}\hspace{-0.2cm} 
\nabla^2 g_i(\boldsymbol{\xi})\vert_{\boldsymbol{\xi}=\boldsymbol{\xi}^{(0)}}=
\begin{bmatrix}
\underbrace{\frac{\partial^2}{\partial \boldsymbol{\theta} \partial \boldsymbol{\theta}^{\top}} g_i(\boldsymbol{\xi})}_{K\times K} & \underbrace{\frac{\partial^2}{\partial \boldsymbol{\theta} \partial \breve{\boldsymbol{\beta}}^{\top}} g_i(\boldsymbol{\xi})}_{K \times (p+1)} &
\underbrace{\frac{\partial^2}{\partial \boldsymbol{\theta} \partial \boldsymbol{\gamma}^{\top}} g_i(\boldsymbol{\xi})}_{K \times q} \\
\\
\underbrace{\frac{\partial^2}{\partial \breve{\boldsymbol{\beta}} \partial \boldsymbol{\theta}^{\top}} g_i(\boldsymbol{\xi})}_{(p+1) \times K} & \underbrace{\frac{\partial^2}{\partial \breve{\boldsymbol{\beta}} \partial \breve{\boldsymbol{\beta}}^{\top}} g_i(\boldsymbol{\xi})}_{(p+1) \times (p+1)} &
\underbrace{\frac{\partial^2}{\partial \breve{\boldsymbol{\beta}} \partial \boldsymbol{\gamma}^{\top}} g_i(\boldsymbol{\xi})}_{(p+1) \times q}\\
\\
\underbrace{\frac{\partial^2}{\partial \boldsymbol{\gamma} \partial \boldsymbol{\theta}^{\top}} g_i(\boldsymbol{\xi})}_{q \times K} & \underbrace{\frac{\partial^2}{\partial \boldsymbol{\gamma} \partial \breve{\boldsymbol{\beta}}^{\top}} g_i(\boldsymbol{\xi})}_{q \times (p+1)} &
\underbrace{\frac{\partial^2}{\partial \boldsymbol{\gamma} \partial \boldsymbol{\gamma}^{\top}} g_i(\boldsymbol{\xi})}_{q \times q}
\end{bmatrix}_{\boldsymbol{\xi}=\boldsymbol{\xi}^{(0)}}.\nonumber
\end{eqnarray}

\newpage 

\noindent To simplify the notation, we define the following quantities:

\vspace{-1cm} 

\begin{eqnarray}
\omega_{0i}&:=& \sum_{j=1}^{j(t_i)} h_0(s_j) \Delta, \nonumber \\
\omega_{0i}^{k}&:=& \sum_{j=1}^{j(t_i)} h_0(s_j) b_k(s_j) \Delta, \nonumber \\
\omega_{0i}^{kl}&:=& \sum_{j=1}^{j(t_i)} h_0(s_j) b_k(s_j) b_l(s_j) \Delta. \nonumber
\end{eqnarray}

\noindent{\textbf{Gradient}}\\

\vspace{-1cm} 

\noindent We first derive the gradient and start with the partial derivatives with respect to the spline coefficients:

\vspace{-1cm}  

\begin{eqnarray}
\frac{\partial}{\partial \theta_k} g_i(\boldsymbol{\xi})&=&\tau_i \left(b_k(t_i)-\exp(\zbf_i^{\top} \bgamma) \sum_{j=1}^{j(t_i)} h_0(s_j) b_k(s_j) \Delta \right)+\frac{(1-\tau_i)}{S_p(t_i \vert \xbf_i, \zbf_i)} \frac{\partial}{\partial \theta_k} S_p(t_i \vert \xbf_i, \zbf_i) \nonumber
\end{eqnarray}

\noindent Note that:

\vspace{-1cm} 

\begin{eqnarray}
\frac{\partial}{\partial \theta_k} S_p(t_i \vert \xbf_i, \zbf_i)=-p(\xbf_i) \exp\left(\zbf_i^{\top} \bgamma-\exp(\zbf_i^{\top} \bgamma) \omega_{0i} \right) \omega_{0i}^k. \nonumber
\end{eqnarray}

\vspace{-0.2cm} 

\noindent It follows that:

\vspace{-1cm} 

\begin{eqnarray} 
\frac{\partial}{\partial \theta_k} g_i(\boldsymbol{\xi})=\tau_i \left(b_k(t_i)-\exp(\zbf_i^{\top} \bgamma) \omega_{0i}^k \right)-\frac{(1-\tau_i)}{S_p(t_i \vert \xbf_i, \zbf_i)} p(\xbf_i) \exp\left(\zbf_i^{\top} \bgamma-\exp(\zbf_i^{\top} \bgamma) \omega_{0i} \right) \omega_{0i}^k,\ k=1,\dots,K. \nonumber 
\end{eqnarray}

\noindent To obtain the derivatives with respect to the $\breve{\bbeta}$ coefficients, let us first compute:

\vspace{-0.8cm}

\begin{eqnarray}
\frac{\partial}{\partial \beta_m} p(\xbf_i)&=& \frac{\partial}{\partial \beta_m} \frac{\exp(\beta_0+\xbf_i^{\top} \bbeta)}{1+\exp(\beta_0+\xbf_i^{\top} \bbeta)} \nonumber \\
&=& \frac{\partial}{\partial \beta_m} \frac{1}{1+\exp(-\beta_0-\xbf_i^{\top} \bbeta)} \nonumber \\
&=& x_{im}(1+\exp(-\beta_0-\xbf_i^{\top} \bbeta))^{-2} \exp(-\beta_0-\xbf_i^{\top} \bbeta) \nonumber \\
&=& \frac{x_{im} \exp(-\beta_0-\xbf_i^{\top} \bbeta)}{\left(\frac{1+\exp(\beta_0+\xbf_i^{\top} \bbeta)}{\exp(\beta_0+\xbf_i^{\top} \bbeta)}\right)^2} \nonumber \\
&=& x_{im} \frac{\exp(\beta_0+\xbf_i^{\top} \bbeta)}{1+\exp(\beta_0+\xbf_i^{\top} \bbeta)} \frac{1}{1+\exp(\beta_0+\xbf_i^{\top} \bbeta)} \nonumber \\ 
&=& x_{im} p(\xbf_i) (1-p(\xbf_i)). \nonumber 
\end{eqnarray}

\newpage 

\noindent It follows that:

\vspace{-1cm} 

\begin{eqnarray}
\frac{\partial}{\partial \beta_m} g_i(\boldsymbol{\xi})&=&\tau_i x_{im} (1-p(\xbf_i))+\frac{(1-\tau_i)}{S_p(t_i \vert \xbf_i, \zbf_i)} x_{im} p(\xbf_i) (1-p(\xbf_i)) \left(\exp(-\exp(\zbf_i^{\top} \bgamma) \omega_{0i})-1\right) \nonumber \\
&& \text{for}\ m=0,\dots,p\ \text{with}\ x_{i0}=1. \nonumber 
\end{eqnarray}

\vspace{-0.3cm} 

\noindent Derivatives with respect to the $\bgamma$ coefficients are:

\vspace{-1cm} 

\begin{eqnarray}
\frac{\partial}{\partial \gamma_s} g_i(\boldsymbol{\xi})= \tau_i z_{is} \left(1-\exp(\zbf_i^{\top} \bgamma) \omega_{0i}\right)-\frac{(1-\tau_i)}{S_p(t_i \vert \xbf_i, \zbf_i)} p(\xbf_i) \exp(\zbf_i^{\top} \bgamma-\exp(\zbf_i^{\top} \bgamma) \omega_{0i}) \omega_{0i} z_{is},\ s=1,\dots,q. \nonumber
\end{eqnarray}

\noindent{\textbf{Hessian}}\\

\vspace{-1cm} 

\noindent To compute the Hessian, we only require the main diagonal and the upper triangular parts (Blocks 12, 13 and 23 below). The lower triangular part is obtained by symmetry.

\vspace{-0.5cm}

\begin{eqnarray}
Block11&:&  \frac{\partial^2}{\partial \theta_k \partial \theta_l} g_i(\boldsymbol{\xi})\hspace{0.4cm} k=1,\dots,K \ \ l=1,\dots,K.\nonumber\\
Block12&:& \frac{\partial^2}{\partial \theta_k \partial \beta_m} g_i(\boldsymbol{\xi})\hspace{0.2cm} k=1,\dots,K \ \ m=0,\dots,p.\nonumber\\
Block13&:& \frac{\partial^2}{\partial \theta_k \partial \gamma_s} g_i(\boldsymbol{\xi})\hspace{0.4cm} k=1,\dots,K \ \ s=1,\dots,q.\nonumber\\
Block22&:&  \frac{\partial^2}{\partial \beta_m \partial \beta_l} g_i(\boldsymbol{\xi})\hspace{0.3cm} l=0,\dots,p \ \ m=0,\dots,p.\nonumber\\
Block23&:& \frac{\partial^2}{\partial \beta_m \partial \gamma_s} g_i(\boldsymbol{\xi})\ \ m=0,\dots,p \ \ s=1,\dots,q.\nonumber\\
Block33&:&  \frac{\partial^2}{\partial \gamma_s \partial \gamma_v} g_i(\boldsymbol{\xi})\hspace{0.4cm} s=1,\dots,q \ \ v=1,\dots,q.\nonumber
\end{eqnarray}

\noindent \underline{\textbf{Block11}}

\noindent Let us first define the function:

\vspace{-1cm} 

\begin{eqnarray}
f_i(\bxi):=\exp(\zbf_i^{\top} \bgamma-\exp(\zbf_i^{\top} \bgamma) \omega_{0i}) \omega_{0i}^k \nonumber 
\end{eqnarray}

\vspace{-0.3cm} 

\noindent It follows that:

\vspace{-1cm} 

\begin{eqnarray}
\frac{\partial^2}{\partial \theta_k \partial \theta_l} g_i(\boldsymbol{\xi})&=&-\tau_i \exp(\zbf_i^{\top} \bgamma) \omega_{0i}^{kl}- (1-\tau_i)p(\xbf_i) \Bigg[\frac{\partial f_i(\bxi)}{\partial \theta_l} S_p(t_i\vert \xbf_i,\zbf_i)-f_i(\bxi) \frac{\partial S_p(t_i\vert \xbf_i,\zbf_i)}{\partial \theta_l}\Bigg]S_p^{-2}(t_i\vert \xbf_i,\zbf_i) \nonumber \\
&& k,l=1,\dots,K,
\end{eqnarray}

\newpage 

\noindent where 

\vspace{-1cm}

\begin{eqnarray}
\frac{\partial f_i(\bxi)}{\partial \theta_l}&=&-\exp(\zbf_i^{\top} \bgamma-\exp(\zbf_i^{\top} \bgamma) \omega_{0i})\exp(\zbf_i^{\top} \bgamma) \omega_{0i}^l \omega_{0i}^k + \exp(\zbf_i^{\top} \bgamma-\exp(\zbf_i^{\top} \bgamma) \omega_{0i}) \omega_{0i}^{kl} \nonumber \\
&=& \exp(\zbf_i^{\top} \bgamma-\exp(\zbf_i^{\top} \bgamma) \omega_{0i}) \omega_{0i}^{kl}-f_i(\bxi) \exp(\zbf_i^{\top} \bgamma) \omega_{0i}^l. \nonumber 
\end{eqnarray}

\vspace{-0.5cm} 

\noindent and

\vspace{-1cm}

\begin{eqnarray}
\frac{\partial}{\partial \theta_l} S_p(t_i \vert \xbf_i, \zbf_i)=-p(\xbf_i) \exp\left(\zbf_i^{\top} \bgamma-\exp(\zbf_i^{\top} \bgamma) \omega_{0i} \right) \omega_{0i}^l. \nonumber
\end{eqnarray}

\noindent \underline{\textbf{Block12}}

\vspace{-1cm}

\begin{eqnarray}
\frac{\partial^2}{\partial \theta_k \partial \beta_m} g_i(\boldsymbol{\xi})&=&-(1-\tau_i) f_i(\bxi) \Bigg[\frac{\partial p(\xbf_i)}{\partial \beta_m} S_p(t_i \vert \xbf_i, \zbf_i)- p(\xbf_i) \frac{\partial S_p(t_i \vert \xbf_i, \zbf_i)}{\partial \beta_m}\Bigg] S_p^{-2}(t_i \vert \xbf_i, \zbf_i) \nonumber \\
&& k=1,\dots,K,\ m=0,\dots,p, \nonumber 
\end{eqnarray}

\vspace{-0.5cm} 

\noindent with 

\vspace{-1cm}

\begin{eqnarray}
\frac{\partial S_p(t_i \vert \xbf_i, \zbf_i)}{\partial \beta_m}= x_{im} p(\xbf_i)(1-p(\xbf_i)) \left(\exp(-\exp(\zbf_i^{\top}\bgamma) \omega_{0i})-1\right) \nonumber 
\end{eqnarray}

\noindent \underline{\textbf{Block13}}

\vspace{-1cm}

\begin{eqnarray}
\frac{\partial^2}{\partial \theta_k \partial \gamma_s} g_i(\boldsymbol{\xi})&=&-\tau_i \exp(\zbf_i^{\top} \bgamma) z_{is} \omega_{0i}^k-(1-\tau_i) p(\xbf_i) \Bigg[\frac{\partial f_i(\bxi)}{\partial \gamma_s} S_p(t_i \vert \xbf_i, \zbf_i)- f_i(\bxi) \frac{\partial S_p(t_i \vert \xbf_i, \zbf_i)}{\partial \gamma_s}\Bigg] S_p^{-2}(t_i \vert \xbf_i, \zbf_i) \nonumber \\
&& k=1,\dots,K,\ s=1,\dots,q, \nonumber 
\end{eqnarray}

\noindent with 

\vspace{-1cm}

\begin{eqnarray}
\frac{\partial S_p(t_i \vert \xbf_i, \zbf_i)}{\partial \gamma_s}&=&- p(\xbf_i)\exp(\zbf_i^{\top}\bgamma-\exp(\zbf_i^{\top}\bgamma) \omega_{0i})z_{is} \omega_{0i}. \nonumber \\
\frac{\partial f_i(\bxi)}{\partial \gamma_s}&=&f_i(\bxi)(z_{is}-\exp(\zbf_i^{\top} \bgamma)z_{is} \omega_{0i}). \nonumber 
\end{eqnarray}

\noindent \underline{\textbf{Block22}}

\noindent Define the following function:

\vspace{-1cm}

\begin{eqnarray}
\widetilde{f}_i(\bxi)=p(\xbf_i)(1-p(\xbf_i)) \left(\exp(-\exp(\zbf_i^{\top}\bgamma)\omega_{0i})-1\right). \nonumber 
\end{eqnarray}

\noindent The second-order derivatives are:

\vspace{-1cm}

\begin{eqnarray}
\frac{\partial^2}{\partial \beta_m \partial \beta_l} g_i(\boldsymbol{\xi})&=&-\tau_i x_{im}x_{il}p(\xbf_i)(1-p(\xbf_i)) \nonumber \\
&&+(1-\tau_i) x_{im} \Bigg[\frac{\partial \widetilde{f}_i(\bxi)}{\partial \beta_l} S_p(t_i \vert \xbf_i, \zbf_i)- \widetilde{f}_i(\bxi) \frac{\partial S_p(t_i \vert \xbf_i, \zbf_i)}{\partial \beta_l}\Bigg]S_p^{-2}(t_i \vert \xbf_i, \zbf_i) \nonumber \\
&&m,l=0,\dots,p, \nonumber 
\end{eqnarray}

\noindent with 

\vspace{-1cm}

\begin{eqnarray}
\frac{\partial \widetilde{f}_i(\bxi)}{\partial \beta_l}&=&x_{il} p(\xbf_i)(1-p(\xbf_i))(1-2p(\xbf_i)) \left(\exp(-\exp(\zbf_i^{\top} \bgamma) \omega_{0i})-1\right).  \nonumber \\ 
\frac{\partial S_p(t_i \vert \xbf_i, \zbf_i)}{\partial \beta_l}&=&x_{il} \widetilde{f}_i(\bxi). \nonumber 
\end{eqnarray} 

\noindent \underline{\textbf{Block23}}

\noindent Define the following function:

\vspace{-1cm}

\begin{eqnarray}
\breve{f}_i(\bxi)=\exp(-\exp(\zbf_i^{\top}\bgamma)\omega_{0i})-1 \nonumber 
\end{eqnarray}

\noindent The second-order partial derivative is:

\vspace{-1cm}

\begin{eqnarray}
\frac{\partial^2}{\partial \beta_m \partial \gamma_s} g_i(\bxi)&=&(1-\tau_i) x_{im} p(\xbf_i)(1-p(\xbf_i)) \Bigg[\frac{\partial \breve{f}_i(\bxi)}{\partial \gamma_s} S_p(t_i \vert \xbf_i, \zbf_i)- \breve{f}_i(\bxi) \frac{\partial S_p(t_i \vert \xbf_i, \zbf_i)}{\partial \gamma_s} \Bigg]S_p^{-2}(t_i \vert \xbf_i, \zbf_i) \nonumber \\
&& m=0\dots,p, \ s=1,\dots,q, \nonumber 
\end{eqnarray}

\vspace{-0.5cm}

\noindent with 

\vspace{-1cm}

\begin{eqnarray}
\frac{\partial \breve{f}_i(\bxi)}{\partial \gamma_s}&=&-\exp(\zbf_i^{\top} \bgamma-\exp(\zbf_i^{\top} \bgamma)\omega_{0i})z_{is} \omega_{0i}. \nonumber 
\end{eqnarray} 

\noindent \underline{\textbf{Block33}}

\noindent Define the following function:

\vspace{-1cm}

\begin{eqnarray}
\ddot{f}_i(\bxi)=\exp(\zbf_i^{\top}\bgamma-\exp(\zbf_i^{\top}\bgamma)\omega_{0i}) \nonumber 
\end{eqnarray}

\noindent The second-order partial derivative is:

\begin{eqnarray}
\frac{\partial^2}{\partial \gamma_s \partial \gamma_v} g_i(\bxi)&=&-\tau_i z_{is} z_{iv} \exp(\zbf_i^{\top} \bgamma)\omega_{0i} \nonumber \\
&&-(1-\tau_i) p(\xbf_i) \omega_{0i} z_{is} \Bigg[\frac{\partial \ddot{f}_i(\bxi)}{\partial \gamma_v} S_p(t_i \vert \xbf_i, \zbf_i)- \ddot{f}_i(\bxi) \frac{\partial S_p(t_i \vert \xbf_i, \zbf_i)}{\partial \gamma_v}\Bigg]S_p^{-2}(t_i \vert \xbf_i, \zbf_i) \nonumber \\
&& s,v=1,\dots,q, \nonumber 
\end{eqnarray}

\noindent with 

\vspace{-1cm}

\begin{eqnarray}
\frac{\partial \ddot{f}_i(\bxi)}{\partial \gamma_v}&=&\ddot{f}_i(\bxi)z_{iv}(1-\exp(\zbf_i^{\top} \bgamma) \omega_{0i}). \nonumber 
\end{eqnarray} 

\noindent \underline{\textbf{Laplace approximation}}

\noindent Define the short notation $\sum_{i=1}^n \gradG:=\nabla g_{\bxi^{(0)}}$ and $\sum_{i=1}^n \hessG:=\nabla^2 g_{\bxi^{(0)}}$ and use \eqref{2Taylor} to write the second-order Taylor expansion of the log-likelihood (omitting constant $c$) as:

\vspace{-1cm} 

\begin{eqnarray} \label{loglik_approx}
\ell(\bxi; \mathcal{D}) &\approx& \sum_{i=1}^n g_i(\bxi) \nonumber\\
&\approx& \bxi^{\top} \left(\nabla g_{\bxi^{(0)}}-\nabla^2 g_{\bxi^{(0)}} \bxi^{(0)}\right)+\frac{1}{2} \bxi^{\top} \nabla^2 g_{\bxi^{(0)}} \bxi.
\end{eqnarray}

\noindent Plugging the approximated log-likelihood \eqref{loglik_approx} into the conditional posterior of the latent vector \eqref{condpost} yields:

\vspace{-1cm}

\begin{eqnarray} \label{Lapprox1}
\widetilde{p}_G(\bxi \vert \lambda, \mathcal{D}) \propto \exp \left( -\frac{1}{2} \bxi^{\top} \big(Q_{\bxi}(\lambda)-\nabla^2 g_{\bxi^{(0)}}\big) \bxi+\bxi^{\top} \left(\nabla g_{\bxi^{(0)}}-\nabla^2 g_{\bxi^{(0)}} \bxi^{(0)}\right) \right). 
\end{eqnarray}

\noindent The logarithm of \eqref{Lapprox1} is thus:

\vspace{-1cm}

\begin{eqnarray} \label{Lapprox2}
\log \widetilde{p}_G(\bxi \vert \lambda, \mathcal{D}) \dot{=} -\frac{1}{2} \bxi^{\top} \big(Q_{\bxi}(\lambda)-\nabla^2 g_{\bxi^{(0)}}\big) \bxi+\bxi^{\top} \left(\nabla g_{\bxi^{(0)}}-\nabla^2 g_{\bxi^{(0)}} \bxi^{(0)}\right),
\end{eqnarray}

\vspace{-0.3cm}

\noindent Taking the gradient of \eqref{Lapprox2} and equating to the zero vector yields:

\vspace{-1cm}

\begin{eqnarray}
\bxi^{(1)}(\lambda)=\big(Q_{\bxi}(\lambda)-\nabla^2 g_{\bxi^{(0)}}\big)^{-1}\left(\nabla g_{\bxi^{(0)}}-\nabla^2 g_{\bxi^{(0)}} \bxi^{(0)}\right). \nonumber 
\end{eqnarray}

\newpage 

\noindent The inverse of the negative Hessian of \eqref{Lapprox2} yields:

\vspace{-1cm}

\begin{eqnarray}
\Sigma_{\bxi}^{(1)}(\lambda)=\big(Q_{\bxi}(\lambda)-\nabla^2 g_{\bxi^{(0)}}\big)^{-1}. \nonumber 
\end{eqnarray}

\vspace{-0.5cm}

\noindent Finally, the Laplace approximation to the conditional posterior latent vector is written as:

\vspace{-1cm} 

\begin{eqnarray}
\widetilde{p}_G(\bxi \vert \lambda, \mathcal{D})=\mathcal{N}_{\text{dim}(\bxi)}\big(\bxi^{(1)}(\lambda), \Sigma_{\bxi}^{(1)}(\lambda)\big). \nonumber 
\end{eqnarray}

\vspace{-0.5cm}

\noindent One can iterate this in a Newton-Raphson type algorithm to obtain a Laplace approximation to the conditional latent vector $\widetilde{p}_G(\bxi \vert \lambda, \mathcal{D})= \mathcal{N}_{\text{dim}(\bxi)}(\bxi^*(\lambda), \Sigma_{\bxi}^*(\lambda))$, where $\bxi^*(\lambda)$ and $\Sigma_{\bxi}^*(\lambda)$ denotes the mean vector and covariance matrix respectively towards which the Newton-Raphson algorithm has converged for a given value of $\lambda$.

\newpage 
\bibliographystyle{apa}
\bibliography{Bibliography_mixture}

\end{document}